\documentclass[twocolumn,prb,aps,english,superscriptaddress]{revtex4}
\usepackage{amsmath}
\usepackage{graphicx}
\usepackage{amssymb}
\usepackage{babel}
\usepackage{color}
\usepackage{hyperref}

\usepackage{mathrsfs}
\usepackage{relsize}

\renewcommand{\v}[1]{|{#1}\rangle}
\newcommand{\iv}[1]{\langle{#1}|}
\renewcommand{\u}{\uparrow}
\renewcommand{\d}{\downarrow}
\newcommand{\U}{\Uparrow}
\newcommand{\D}{\Downarrow}

\newcommand{\sech}{{\rm \,sech}}

\newcommand{\N}{{\cal N}}

\newcommand{\I}{{\rm I}}
\newcommand{\II}{{\rm II}}
\newcommand{\LI}[2]{{\Lambda}^{(#1)}_{\I,#2}}
\newcommand{\LpI}[2]{{\bar\Lambda}^{(#1)}_{\I,#2}}
\newcommand{\LII}[2]{{\Lambda}^{(#1)}_{\II,#2}}

\newcommand{\lI}[1]{{\Lambda}^{(#1)}_{\I}}
\newcommand{\lpI}[1]{{\bar\Lambda}^{(#1)}_{\I}}
\newcommand{\lII}[1]{{\Lambda}^{(#1)}_{\II}}

\newcommand{\is}{{\frak i}}
\newcommand{\fs}{{\frak f}}

\renewcommand{\O}{{\cal O}}

\begin{document}

\author{Dmitry Solenov}
\email{d.solenov@gmail.com}
\altaffiliation{Present address: Naval Research Laboratory, 4555 Overlook Ave., SW Washington, District of Columbia 20375, USA}
\affiliation{National Research Council, National Academies, Washington, District of Columbia 20001, USA}

\author{Sophia E. Economou}
\affiliation{Naval Research Laboratory, Washington, District of
Columbia 20375, USA}
\author{Thomas L. Reinecke}
\affiliation{Naval Research Laboratory, Washington, District of
Columbia 20375, USA}

\title{Excitation spectrum as a resource for efficient two-qubit entangling gates}

\begin{abstract}
Physical systems representing qubits typically have one or more accessible quantum states in addition to the two states that encode the qubit. We demonstrate that active involvement of such auxiliary states can be beneficial in constructing entangling two-qubit operations. We investigate the general case of two multi-state quantum systems coupled via a quantum resonator. The approach is illustrated with the examples of three systems: self-assembled InAs/GaAs quantum dots, NV-centers in diamond, and superconducting transmon qubits. Fidelities of the gate operations are calculated based on numerical simulations of each system.
\end{abstract}
\maketitle

\section{introduction}

Quantum information has been traditionally built around two-state quantum systems representing quantum bits (qubits).\cite{nielsenchuang} In order to perform operations on a single or multiple qubits, qubit states must interact with control fields and with the fields that mediate the interaction. The control is typically done by a classical (coherent state) bosonic field, such as laser or microwave field that drives the evolution of the system.\cite{Kennedy,Awschalom-SiC,Carter} Classical control, however, is insufficient to generate entanglement. The qubit states must interact with one another via a quantum field, such as an optical or a microwave cavity mode. Thus, quantum two-qubit operations result from a combination of classical control and (quantum) physical interaction between the qubits.\cite{Kim,Lucero,nielsenchuang}

Traditionally, in many quantum computing systems\cite{LaddJelezko,Buluta} the focus was on either coupling qubit transitions directly to the cavity mode by tuning them into the resonance or by coupling the qubits indirectly via a third state ($\Lambda$-system), eliminating the latter from the evolution.\cite{Lucero,Nori,nielsenchuang} In both cases, the system is carried through a resonance condition either by driving the system with external control fields or by slow modification of system parameters with time.

\begin{figure}
\includegraphics[width=0.99\columnwidth]{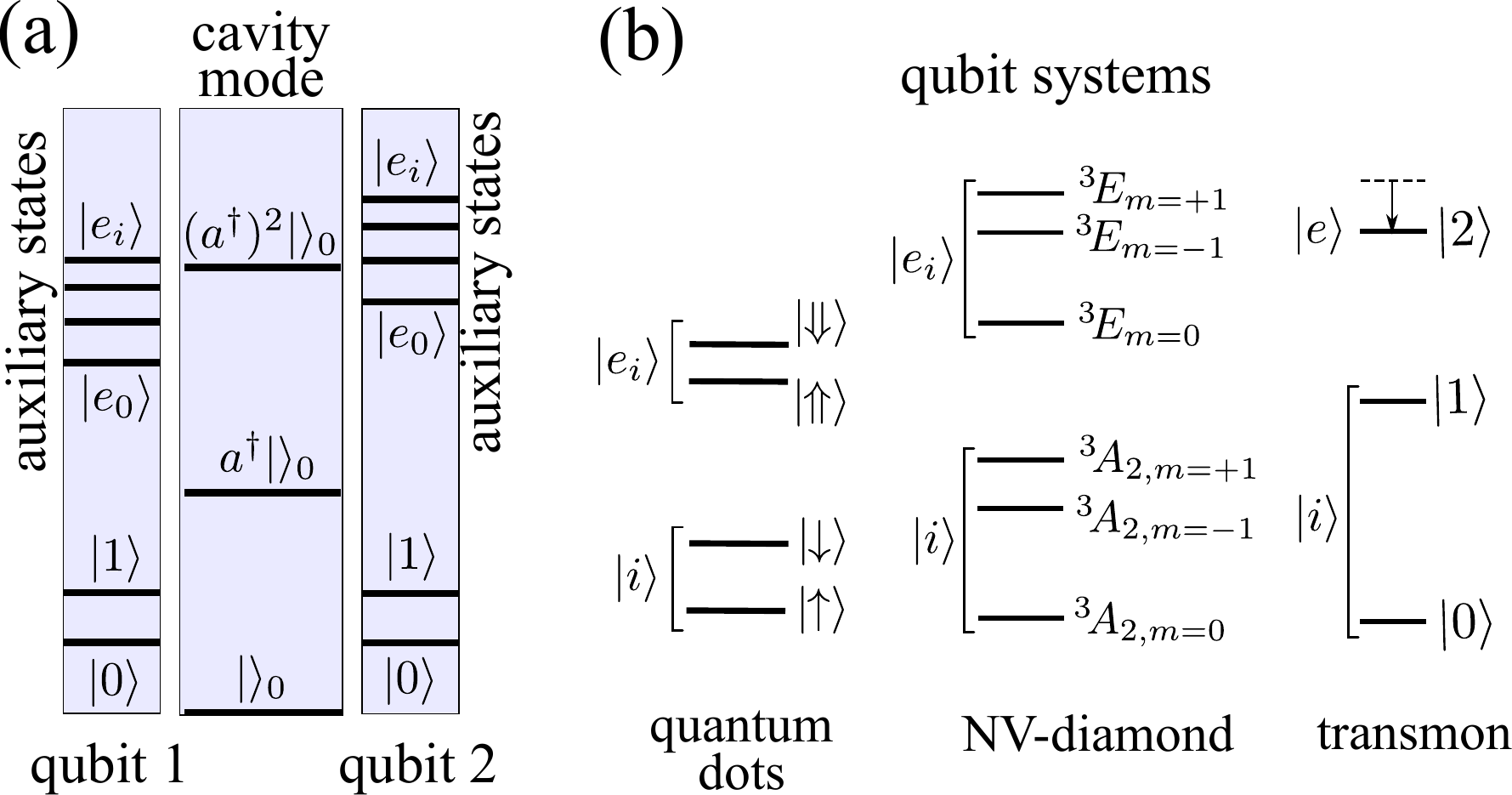}
\caption{\label{fig:2QC-cartoon}
The qubit-cavity system.
(a) Two multistate qubit systems interacting via the cavity mode. (b) Different physical systems considered here as examples: self-assembled quantum dots (charge exciton states as $\v{e_i}$), NV-centers in diamond (only the first two states of set $\v{i}$ are used to encode a qubit), transmon superconducting qubits (only the first three states are shown here).
}
\end{figure}

A qubit system typically has multiple distinct quantum states above the two states that are used to encode the qubit. These states can also be used to manipulate qubits performing single-qubit gates (e.g., via a $\Lambda$-system \cite{Awschalom-SiC,Carter}) or two-qubit gates \cite{Strauch,Matsuo,Lu-Nori,transmon}. Unlike earlier two-qubit gate schemes, in which auxiliary states play a role, the approach we use here is based on a regime of operation that is common to most qubit-cavity systems. In this paper we generalize our earlier derivations made for specific systems\cite{solenov-NV,solenov-QD} and show that active control of populations of additional (auxiliary) states can be beneficial in constructing various two-qubit operations. To this end, we demonstrate how our results can be applied to a superconducting transmon-based system\cite{SC-review,transmon} that differs substantially from the optically-controlled systems\cite{solenov-NV,solenov-QD} investigated earlier (see Fig.~\ref{fig:2QC-cartoon}). We focus on two-qubit entangling operations and demonstrate that efficient two-qubit gates can be performed by driving the system along certain trajectories through excited states of the qubit-cavity system using a set of a few simple local pulses. 

While the spectrum of {\it each} individual system representing a qubit is specific to the physics of that system, the spectrum of {\it two} such qubit systems interacting with a cavity mode has a structure that is common to many physical realizations. When the cavity mode is detuned far from transitions in the qubit systems, it cannot effectively mediate an interaction, and the qubits remain isolated from one another. In this case only single-qubit operations are possible. When the cavity mode frequency is in resonance with transitions in both qubit systems, non-local (entangled) superpositions of states form and entanglement can be manipulated. At that time, single-qubit operations become difficult. We show that in the intermediate regime of resonance when the qubits are already sufficiently isolated, the auxiliary excited states can still develop substantial non-local superpositions, which can be used to perform two-qubit entangling operations.

The organization of the paper is as follows: in Sec. \ref{sec:2QC} we introduce the basic structure of a pair of multi-state qubit systems coupled to a single cavity mode. In Sec. \ref{sec:1Q} we outline the basic concepts involved in typical deterministic single-qubit operations performed via auxiliary states or directly. In Sec. \ref{sec:2Q} we develop a generalized approach for two-qubit gates performed via auxiliary states. In the remaining three sections we give three specific examples [see Fig.~\ref{fig:2QC-cartoon}(b)]: (i) self-assembled  quantum dots, Sec.~\ref{sec:QD}; (ii) NV-centers in diamond, Sec.~\ref{sec:NV}; and (iii) superconducting (transmon) qubits, Sec.~\ref{sec:SC}. In these sections we discuss details specific to each system and perform numerical simulations of two-qubit operations in realistic conditions. In particular, in Secs. \ref{sec:QD} and \ref{sec:NV} we review our earlier results on entangling gates,\cite{solenov-QD,solenov-NV} and in Sec.~\ref{sec:SC} we apply our approach to a superconducting transmon system. Fidelities of the gate operations are numerically analyzed for all systems. 

\section{Two qubits and a cavity}
\label{sec:2QC}

\subsection{states and coherent evolution}

Two multi-state qubit systems, each interacting with photons [see Fig.~\ref{fig:2QC-cartoon}(a)], can be described by standard light-matter interaction Hamiltonian,\cite{Louisell,Mandel}
\begin{eqnarray}\label{eq:2QC:H-full}
&&H = \sum_{\xi\zeta}(\varepsilon^{(1)}_{\xi}+\varepsilon^{(2)}_{\zeta}) \v{\xi\zeta}\iv{\xi\zeta} 
+ \sum_{s}\omega_s a_s^\dag a_s
\\\nonumber
&&+
\sum_{\xi\xi'\zeta s}\!
(a_s\!\!+\!a_s^\dag)\left( 
{\frak g}^{(1)}_{s,\xi\xi'}\v{\xi\zeta}\iv{\xi'\zeta}
+
{\frak g}^{(2)}_{s,\xi\xi'}\v{\zeta\xi}\iv{\zeta\xi'}
+h.c.
\right),
\end{eqnarray}
where the first index ($\xi$) in $\v{\xi\zeta}$ is the state of the first multi-state qubit (qubit-1) system and the second index ($\zeta$) corresponds to the state of the second multi-state qubit (qubit-2) system. One of the bosonic modes ($s=0$) represents the cavity mode with frequency $\omega_C$, and the other bosonic fields ($s>0$) are in the coherent state\cite{Louisell} and represent the classical time-dependent pulse control part of the Hamiltonian, $V(t)$. We will omit index $s=0$ for the cavity mode operators from now on to shorten notations.

We will use the first two states in each qubit system $\v{i=0,1}$ to represent a qubit. Higher energy auxiliary states will be denoted by $\v{e_{i=0,1,...}}$. Here we will focus on the system in which photons couple only to transitions between $\v{i}$ and $\v{e_j}$ in each qubit system. This assumption allows for simpler derivations, and we will keep it for clarity of presentation. It can later be relaxed as we demonstrate in Sec.~\ref{sec:SC}. For many physical realizations of qubit systems it is also sufficient to treat the interaction with photons within the rotating wave approximation.\cite{Louisell,Mandel,Abragam} With these assumptions, the Hamiltonian can be formulated in a more convenient form:
\begin{eqnarray}\label{eq:2QC:H}
H = H_0 + V(t),
\end{eqnarray}
with
\begin{eqnarray}\nonumber
&&H_0 = 
\sum_{ij}\!\varepsilon_{ij}\v{ij}\iv{ij} 
+
\sum_{ij}\!
\left(\!
\varepsilon^{(1)}_{ij}\v{e_ij}\iv{e_ij} 
\!+\!
\varepsilon^{(2)}_{ij}\v{ie_j}\iv{ie_j} 
\right)
\\\label{eq:2QC:H0}
&&
+
\sum_{ij}
\varepsilon^{ee}_{ij}\v{e_ie_j}\iv{e_ie_j} 
+
\omega_C a^\dag a
\\\nonumber
&&+
\sum_{i>j,\xi}\!
\left( 
g^{(1)}_{ij}a\v{e_i\xi}\iv{j\xi}
+
g^{(2)}_{ij}a\v{\xi e_i}\iv{\xi j}
+h.c.
\right),
\end{eqnarray}
and
\begin{eqnarray}\label{eq:2QC:V}
V(t)&=& \sum_p \Omega_p(t-t_p)\cos(\omega_pt+\phi_p)
\\\nonumber
&\times&
\sum_{ij;\xi} 
\left(
{\frak u}^{(1)}_{ij} \v{e_i\xi}\iv{j\xi}
+{\frak u}^{(2)}_{ij} \v{\xi e_i}\iv{\xi j} 
+h.c. 
\right)
\!,
\end{eqnarray}
where $\xi$ denotes qubit ($i$) and auxiliary ($e_i$) states as before, constants ${\frak u}_{ij}^{(n)}$ define allowed transitions, and $\omega_p$, $\phi_p$, and $\Omega_p$ are the frequency, phase, and shape of the $p$-th pulse respectively. Specific physical realizations may also include states that are in the same range of energies as states $\v{i}$ and $\v{e_i}$, but do not participate in the coherent evolution discussed in this work. Some of these states can participate in incoherent dynamics (dissipation) as, e.g., in the case of NV-centers in diamond discussed in Sec.~\ref{sec:NV}. We will include these states in numerical simulations, but omit them in the discussion of coherent dynamics. The combined energies of the two-qubit system are defined as $\varepsilon_{ij} = \varepsilon^{(1)}_{i}+\varepsilon^{(2)}_{j}$, $\varepsilon^{(1)}_{ij} = \epsilon^{(1)}_{e_i}+\varepsilon^{(2)}_{j}$, $\varepsilon^{(2)}_{ij} = \varepsilon^{(1)}_{i}+\varepsilon^{(2)}_{e_j}$, and $\varepsilon^{ee}_{ij} = \varepsilon^{(1)}_{e_i}+\varepsilon^{(2)}_{e_j}$. We will assume that $\Delta=(\varepsilon^{(2)}_{e_i}-\varepsilon^{(1)}_{e_i})-(\varepsilon^{(2)}_{i}-\varepsilon^{(1)}_{i})>0$, and $g^{(n)}_{ij}=g>0$. We will drop the dependence of $\Delta$ and $g$ (for allowed transitions) on the state indexes, assuming that the variation are small. The results will not change qualitatively if this dependence is restored. The complex argument of $g$ can also be ignored unless indicated otherwise.

\begin{figure}
\includegraphics[width=0.8\columnwidth]{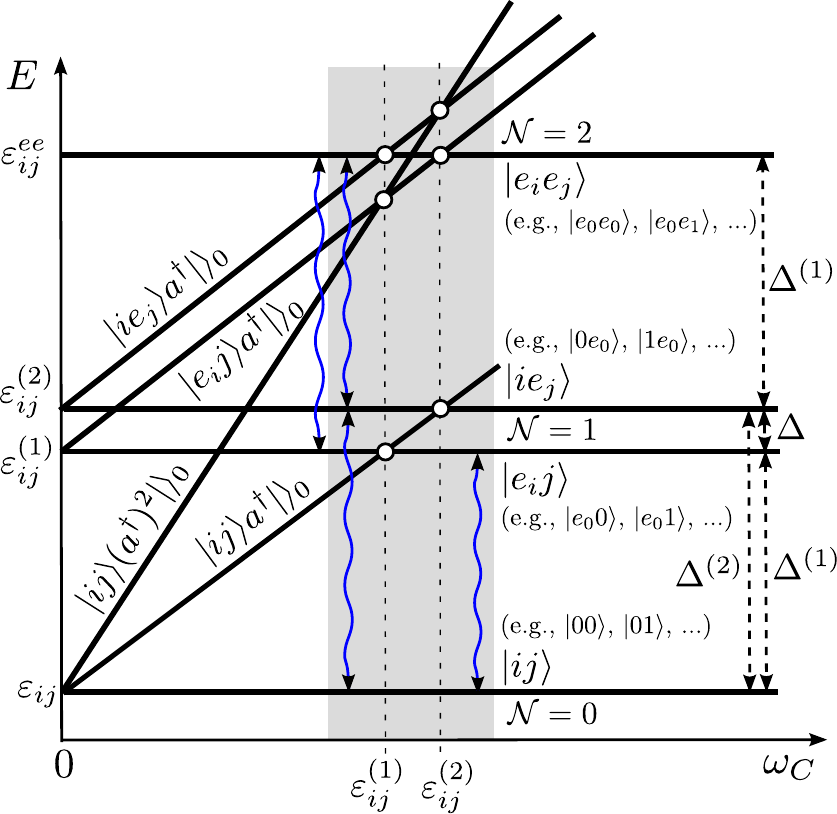}
\caption{\label{fig:2QC:spec}
Schematic depiction of the spectrum of $H_0$ as a function of the cavity mode frequency $\omega_C$. Auxiliary excited states form several bands that anticross with the qubit-photon bands. The anticrossings are shown schematically as circled white dots. The range of $\omega_C$ relevant for pulse-controlled single- and two-qubit operations is highlighted by shaded area. Wavy lines indicate transitions that can be excited by classical control fields (similar transitions between photon bands are not shown).}
\end{figure}

The spectrum of the time-independent part of the Hamiltonian
\begin{eqnarray}\label{eq:2QC:H0-spectrum}
H_0\v{\psi_i}=E_i\v{\psi_i}
\end{eqnarray}
is shown schematically in Fig.~\ref{fig:2QC:spec} as a function of the cavity mode frequency $\omega_C$. In our case it is convenient to focus on the total energy spectrum and not the quasiparticle spectrum.\cite{solenov-QD,solenov-NV,Mahan} Figure~\ref{fig:2QC:spec} also outlines the notations used in $H_0$. With the above assumptions the Hamiltonian $H_0$ conserves the total number of excitations in the system:
\begin{eqnarray}\label{eq:2QC:N}
\N = \sum_{i\xi}(
\v{e_i\xi}\iv{e_i\xi} + \v{\xi e_i}\iv{\xi e_i}
)
+ a^\dag a.
\end{eqnarray}
As a result, the spectrum (total energy) can be separated into groups of states each having a particular total number of excitations $\N$. Note that we use a non-standard convention for counting excitations\cite{excitations-comment} that is particularly convenient for describing quantum operations that will be introduced in the next sections.

\begin{figure}
\includegraphics[width=0.8\columnwidth]{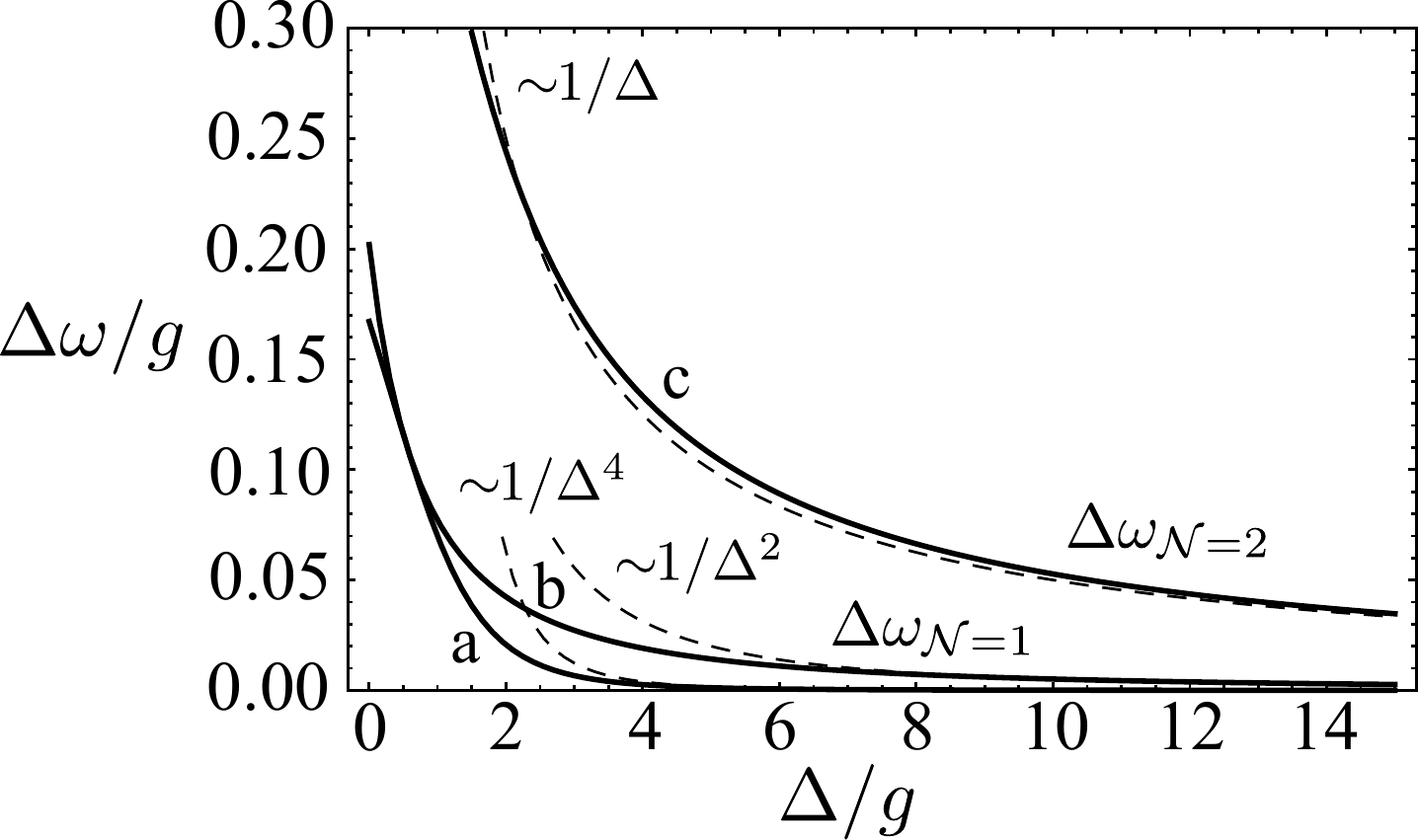}
\caption{\label{fig:2QC:int}
Splitting between transitions due to interaction via the cavity: numerical results and analytical asymptotic behavior. The curves (a) and (b) correspond to $\omega_{\N=1}$ with $\omega_C=\varepsilon^{(2)}$ and $\omega_C=\varepsilon^{(1)}$ respectively. The (c) curve corresponds to $\omega_{\N=2}$ and $\omega_C=\varepsilon^{(2)}$. The dashed curves show asymptotic behavior derived in Appendix~\ref{app:interaction}. The calculations are performed for a system with two auxiliary states $\v{e_0}$ and $\v{e_1}$ for each qubit.
}
\end{figure}

The zero-excitation part of the spectrum, $\N=0$, contains the qubit states $\v{00}$, $\v{01}$, $\v{10}$, and $\v{11}$. This qubit subspace is unaffected by the cavity mode.
The one-excitation part of the spectrum, $\N=1$, can be separated into three bands: the one-photon band, $\v{ij}a^\dag\v{}_0$; qubit-1 band, $\v{e_ij}$; and qubit-2 band, $\v{ie_j}$. Near the anti-crossings where the $\v{ij}a^\dag\v{}_0$ photon band intersects $\v{e_ij}$ or $\v{e_ij}$, these states mix. When $\Delta\gg g$ [see Fig.~\ref{fig:2QC:spec}],
the anti-crossings corresponding to different qubit systems become isolated from one another. Each qubit can still couple strongly to the cavity at some values of $\omega_C$ and form a state $\v{``e_i"}$ that does not separate into a direct product of a qubit-system and a cavity-photon state.
At the same time, the qubit systems do not couple to each other through the cavity. The states $\v{``e_i"j}$ and $\v{i``e_j"}$ remain local and can be used in single qubit manipulations,\cite{Carter} as we will describe in the next section. When $\Delta\ll g$, states $\v{``e_i"j}$ and $\v{i``e_j"}$ mix, forming entangled states that can be used for two-qubit rotations.\cite{economou-2qb, Kim} However the ability to do single qubit operations with simple pulses using these states\cite{Kennedy,Awschalom-SiC,economou-1qb,Yale} is lost. This transition from a strong resonance regime to an off-resonance regime can be described by looking at, e.g., the difference
\begin{eqnarray}\label{eq:2QC:omega-N1}
\Delta\omega_{\N=1} = \omega_{01\leftrightarrow e_01} - \omega_{00\leftrightarrow e_00},
\end{eqnarray}
where $\omega_{\xi\zeta\leftrightarrow\xi'\zeta'}$ is the transition energy between states $\v{``\xi\zeta"}$ and $\v{``\xi'\zeta'"}$ (quotation marks have the same meaning as before). When $\Delta\omega_{\N=1}=0$, the interaction between qubits in the $\N=1$ sector vanishes: transition energies (e.g., $0\leftrightarrow e_0$) in the qubit-1 system are not affected by the state of the second qubit. The value of $\Delta\omega_{\N=1}$ as a function of $\Delta$ is shown in Fig.~\ref{fig:2QC:int} and will be discussed in detail in Sec.~\ref{sec:2Q:inter}.

The two-excitation part of the spectrum, $\N=2$, contains four bands: (i) the two-photon band, (ii) the band with one-photon and one excitation in the qubit-1 system, (iii) the band with one-photon and one excitation in the qubit-2 system, and (iv) the band with two qubit excitations, one in each qubit. These bands intersect within the same range of $\omega_C$ as do the bands in the $\N=1$ sector, but develop more anti-crossing features. As a result, the interaction between qubit systems mediated by cavity photons can be stronger for $\N=2$ states. This can be seen by looking at, e.g., the energy difference
\begin{eqnarray}\label{eq:2QC:omega-N2}
\Delta\omega_{\N=2} = \omega_{10\leftrightarrow 1e_0} - \omega_{e_00\leftrightarrow e_0e_0}.
\end{eqnarray}
This is almost the same as $\Delta\omega_{\N=1}$ formulated for the transitions in the qubit-2 system, except that the qubit-1 system in the second term of Eq.~(\ref{eq:2QC:omega-N2}) is brought up to the $\N=1$ sector and, thus, an excitation in the qubit-2 system brings the entire system into the $\N=2$ sector. The specific choice of states will become apparent in Sec.~\ref{sec:2Q:inter}, where $\Delta\omega_{\N=2}$ is discussed in detail. As before, $\Delta\omega_{\N=2}=0$ means that there is no interaction between the qubit systems. The numerical and analytical results for $\Delta\omega_{\N=2}$ are shown in Fig.~\ref{fig:2QC:int}. Both $\Delta\omega_{\N=1}$ and $\Delta\omega_{\N=2}$ decrease as $\Delta$ becomes larger, and the qubit systems are completely isolated spectrally at $\Delta\to\infty$. At the same time, in many cases $\Delta\omega_{\N=2}$ decreases substantially slower compared to $\Delta\omega_{\N=1}$. This will be the basis for the two-qubit entangling operations developed in Sec.~\ref{sec:2Q}. If the restriction on the cavity-qubit coupling is relaxed, Eq.~(\ref{eq:2QC:N}) that defines $\N$ will change. However, in many cases, the notation introduced above can still be employed. One such example will be discussed in Sec.~\ref{sec:SC}.   

Investigation of the dynamics of the system due to control pulses is best described in the energy basis where the total Hamiltonian takes the form
\begin{eqnarray}\label{eq:2QC:H-diag}
{\cal H}(t) = \sum_{i}E_i \v{\psi_i}\iv{\psi_i} 
+ {\cal V}(t),
\end{eqnarray}
where
\begin{eqnarray}\label{eq:2QC:V-E}
{\cal V}(t\!)\!\!=\!\!\! \sum_p\! \Omega_p(t\!\!-\!t_p)
\!\!\sum_{ij,\pm}\!\!\left(
\!{\frak M}^{\pm\phi_p}_{ij}
\!\v{\psi_i}\!\iv{\psi_j}
e^{i(\omega_{i\!\leftrightarrow\! j}\pm \omega_p\!)t} 
\!\! + \! h.c. 
\!\right)\!\!,
\end{eqnarray}
and
\begin{eqnarray}\label{eq:2QC:Mnij}
{\frak M}^{\phi}_{ij} \!\!=\! 
\frac{e^{i\phi}}{2}
\!\!\!\sum_{nm,\xi}\!\!
\iv{\psi_i}
\!\left(\!
{\frak u}_{nm}^{(1)}\v{n\xi}\!\iv{m\xi}
\!+\!
{\frak u}_{nm}^{(2)}\v{\xi n}\!\iv{\xi m}
\!\right)\!
\v{\psi_j}.
\end{eqnarray}

\subsection{decoherence}

Driving the qubit system with classical control pulses, $\Omega_p(t)$, can become non-coherent in several ways. Typically pulses are designed to address only a few specific transitions. Other transitions that are spectrally close can also partially participate, resulting in incorrect accumulation of phases or even leakage from the qubit subspace. Another significant source of coherence loss is interaction with environmental degrees of freedom. The presence of a large number of states interacting weakly with the system makes the formulation of the dynamics via quantum states intractable, and a reduced density matrix formulation\cite{Mahan,Blum,vanKampen,FeynmanHibbs} has to be employed. The reduced density matrix with all the external degrees of freedom (bath) traced over is defined\cite{vanKampen,FeynmanHibbs,Mahan} as 
\begin{eqnarray}\label{eq:dec:rho}
\rho(t) = {\rm Tr}_B\left( Te^{-i\int_0^t dt{\mathscr H}} \rho_{total}(0) Te^{i\int_0^t dt{\mathscr H}} \right),
\end{eqnarray}
where $T$ is time-ordering operator. The initial total density matrix, $\rho_{total}(0)$, is often assumed to be factorized\cite{vanKampen} $\rho_{total}(0) = \rho(0)\otimes\rho_B$ into the system and environment parts. The total Hamiltonian, $\mathscr H$, of the system and environment is
\begin{eqnarray}\label{eq:dec:H}
{\mathscr H} = H(t) + H_B + H_{SB},
\end{eqnarray}
where $H_B$ describes environmental degrees of freedom and $H_{SB}$ is their interaction with the system. For example, a bosonic environment (e.g., photon or phonons) can be described as\cite{Leggett} $H_B = \sum_{ik} \omega_{ik} b^\dag_{ik} b_{ik}$ and $H_{SB} = \sum_{ik} \left( {\frak g}_{ik} b^\dag_{ik}{\frak O}_i + h.c. \right)$, where ${\frak O}$ is an operator specific to a particular system. In the limit of large times\cite{vanKampen,solenov1,solenov2,solenov2a,solenov2b} ($t\gg 1/{\frak g}$), the effect of the environment can be approximately represented by several constants, and the evolution of the reduced density matrix can be reduced to a simpler master-equation form\cite{vanKampen}
\begin{eqnarray}\label{eq:dec:rho-eq}
i\dot{\rho} &=& [H(t),\rho]+\sum_s i\Gamma_s L_s\{\rho\},
\\\label{eq:dec:L}
L_s\{\rho\}&=&[P_s\rho P_s^\dag - (P^\dag_sP_s\rho+\rho P^\dag_s P_s)/2],
\end{eqnarray}
where $\Gamma_s$ are decay rates for transitions $\v{\xi_s}\to\v{\zeta_s}$, and $P_s = \v{\zeta_s}\iv{\xi_s}$ are the projection operators [$P_s = (a^\dag)^{n}\v{}_{0\,0}\!\iv{}a^{n+1}$ for the cavity mode states]. Numerical simulation of the system is most easily performed in the energy basis, in which Eq.~(\ref{eq:dec:rho-eq}) simplifies to
\begin{eqnarray}\label{eq:dec:rho-E-eq}
i\dot{\rho_E} = [{\cal V}(t),\rho_E]+\sum_s i\Gamma_s {\cal L}_s\{\rho_E\},
\end{eqnarray}
where ${\cal L}\{\rho_E\} = U^\dag_E L\{\rho\} U_E$, $\rho_E = U^\dag_E\rho U_E$, and $U_{E,ij} = \v{\psi_i}\iv{\psi_j}$.

\section{Single qubit gates}
\label{sec:1Q}

Single qubit gates are the basic building blocks for quantum computing. Single-qubit deterministic (reversible) quantum gates are unitary operations, $U$, applied to the qubit.\cite{nielsenchuang} They perform rotations of the qubit basis states (frame of reference). For an arbitrary qubit state $\v{\psi_\is} = C_0\v{0}+C_1\v{1}$ before the gate, the state after the gate is
\begin{eqnarray}\label{eq:1Q:rotate-basis}
\v{\psi_\fs} = U\v{\psi_\is} = C_0 U\v{0}+C_1 U\v{1},
\end{eqnarray}
or, equivalently, $C'_i=U_{ij} C_j$ and $U_{ij} = \iv{i}U\v{j}$.
In order to avoid dealing with qubit precession due to differences in energy between states $\v{0}$ and $\v{1}$, the qubits are typically formed in the rotating frame of reference, so that the total evolution operator is $U_{tot}(t) = e^{-iH_0t}U(t)$. In this case, the gates are due to the evolution operator, $U$, in the interaction representation.   

\begin{figure}
\includegraphics[width=0.4\columnwidth]{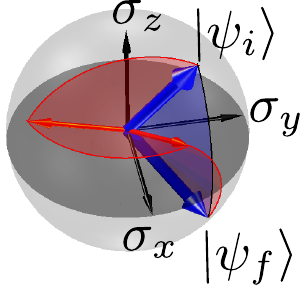}
\caption{\label{fig:1Q:bloch}
Single-qubit gate as rotation of the basis around three orthogonal directions $\sigma_x$, $\sigma_y$, $\sigma_z$.}
\end{figure}

Due to the definition (\ref{eq:1Q:rotate-basis}), reversible quantum gates are often based on classical reversible gates operating on classical bits (0 or 1). For example, the quantum {\it single-qubit} swap gate (typically referred to as the $X$ gate), $C_0\v{0}+C_1\v{1}\to C_0\v{1}+C_1\v{0}$ is based on the classical NOT operation, $0\to 1$ and $1\to 0$. In addition, quantum gates can alter the phase of the qubit amplitudes. For instance, the $Z$ gate adds a phase of $\pi$ to $C_1$, which means that $U_Z = \sigma_z$ (Pauli matrix) acting in the space of $\v{0}$ and $\v{1}$. An arbitrary single-qubit operation can be formulated as a rotation of the basis around three directions ($x$, $y$, $z$)
\begin{eqnarray}\label{eq:1Q:pauli-rotate}
U = \exp[-\theta \vec{v}\cdot\vec{\sigma}/2]
=
T\exp[-i\int_0^tdt'{\cal V}(t')],
\end{eqnarray}
where $\vec{\sigma} = \{\sigma_x,\sigma_y,\sigma_z\}$, $\vec{v}$ is a unit vector that defines the axis of rotation (see Fig.~\ref{fig:1Q:bloch}), and ${\cal V}(t)$ is external classical filed control defined in Eqs.~(\ref{eq:2QC:V}) and (\ref{eq:2QC:V-E}). Note that if the control field is not in resonance with the targeted transition, then $[{\cal V}(t),{\cal V}(t')]\neq 0$ for $t\neq t'$, and the time ordering has to be performed when calculating the exponent. The overall phase accumulated during the evolution is not important and will be omitted, unless stated otherwise.

\begin{figure}
\includegraphics[width=0.5\columnwidth]{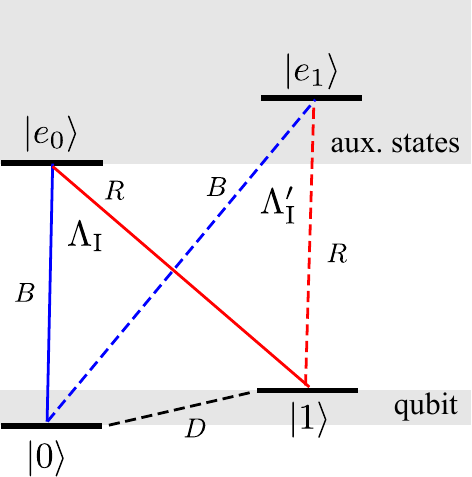}
\caption{\label{fig:1Q:lambdas}
Possible transitions for a single qubit system. Transitions to the first two auxiliary states, $\v{e_0}$ and $\v{e_1}$, are shown as an example. Each auxiliary state forms a $\Lambda$ system ($\Lambda_\I$ and $\Lambda'_\I$ for $\v{e_0}$ and $\v{e_1}$ respectively) in which the higher-frequency transition is denoted as $B$ (blue), and the lower-frequency transition as $R$ (red). The direct transition between the qubit states is shown as $D$.
}
\end{figure}

In a two-qubit system coupled via a cavity mode, single-qubit gates can be performed in several ways. The control field can be coupled directly to the qubit states.\cite{Lucero} For the three systems considered later in Secs.~\ref{sec:QD}, \ref{sec:NV}, and \ref{sec:SC}, this requires $\omega_p$ to be in a microwave range. Often it is more effective for a given realization to perform single-qubit rotations using a higher energy auxiliary state, e.g., $\v{e}$, that together with the states $\v{0}$ and $\v{1}$ forms a $\Lambda$ system\cite{economou-1qb,Carter} (see Fig.~\ref{fig:1Q:lambdas}). We will focus on this case here. In order to be used in a single-qubit manipulation, transitions to the auxiliary state have to be independent of the state of the other qubit:\cite{1qb-transitions-comment,Ashhab} for example, transitions $\v{00}\leftrightarrow\v{``e"0}$ and $\v{01}\leftrightarrow\v{``e"1}$ have to be indistinguishable. This can be measured by the transition energy difference $\Delta\omega_{\N=1}$, introduced in the previous section. Although the cavity mode always interacts with the qubit systems, the value of $\Delta\omega_{\N=1}$ (see Fig.~\ref{fig:2QC:int}) decreases rapidly as $\Delta$ becomes large compared to $g$ (qubit-cavity coupling). This occurs for any value of $\omega_C$, since the cavity cannot be in resonance with the corresponding transitions in the qubit-1 and qubit-2 system at the same time. As a result, for $\Delta\gg g$ the interaction between the qubit systems cannot be effectively mediated by the cavity photons and the state of the other qubit can be factored out with sufficient accuracy (see Appendix~\ref{app:interaction} for details).

Here we provide two examples that give insight into the composition of two-qubit gates in the next section. First, consider a diagonal operation with $\vec{v}=\{0,0,1\}$ in the rotating frame of reference,
\begin{eqnarray}\label{eq:1Q:Z}
U_Z(\phi) = \left(
\begin{array}{cccc}
e^{i\phi} & 0\\
0 & 1\\
\end{array}
\right).
\end{eqnarray}
The necessary phase can be accumulated for state $\v{0}$ (or $\v{1}$) by applying a control pulse\cite{economou-1qb} to transition $\v{0}\leftrightarrow\v{e}$ (transition $\Lambda_{\I,B}$ in Fig.~\ref{fig:1Q:lambdas}). The pulse frequency ($\omega_p$) has to be tuned near the resonance with some detuning $\delta_p = \omega_p -  \omega_{\Lambda_{\I,B}}$, where $\omega_{\Lambda_{\I,B}}=\omega_{00\leftrightarrow e_00}=\omega_{01\leftrightarrow e_01}$ (with acceptable tolerance). The dynamics of such a non-resonant driving cannot be solved analytically in general. However, when $\Omega(t)=\Omega\sech(\sigma t)$ an exact analytical solution is available (Rosen-Zener pulse shape; see Ref.~\onlinecite{rosenzener}): the occupation probability returns to state $\v{0}$ completely ($2\pi$ pulse) when $\Omega = \sigma$. The phase acquired by state $\v{0}$ after the pulse is given by\cite{economou-1qb,rosenzener}
\begin{eqnarray}\label{eq:1Q:Z-phase}
e^{i\phi} = - \frac{\sigma+i\delta_p}{\sigma-i\delta_p}.
\end{eqnarray}
During the gate operation the population of state $\v{0}$ is carried through state $\v{e}$ and returned back with phase factor of $e^{i\phi}$. The case $\phi=\pi$ is achieved when the pulse is in resonance, $\delta_p=0$, and it corresponds to the standard $Z$ gate. Note that the phase of the pulse field $\phi_p$ [see Eqs.~\ref{eq:2QC:V} and \ref{eq:2QC:V-E}] does not affect the gate operation.

The two rotations orthogonal to $\vec{v}=\{0,0,1\}$ can be performed using both $\Lambda_{\I,B}$ and $\Lambda_{\I,R}$ (see  Fig.~\ref{fig:1Q:lambdas}). A two-color pulse with frequencies $\omega_1$ and $\omega_2$ that are off-resonance with both $\omega_{\Lambda_{\I,B}}$ and $\omega_{\Lambda_{\I,R}}$, but
with $\omega_1-\omega_2 = \varepsilon_{10} - \varepsilon_{00}$, can be used in this case,\cite{Carter,Yale} resulting in the gate evolution operator
\begin{eqnarray}\label{eq:1Q:XY}
U_{XY}(\theta,\phi) = \exp[i\theta (\v{0}\iv{1}e^{i\phi}+\v{1}\iv{0}e^{-i\phi})/2],
\end{eqnarray}
where, $\phi = \phi_2-\phi_1$ [see Eqs.~\ref{eq:2QC:V} and \ref{eq:2QC:V-E}] and $\theta=2|{\frak u}|\int_{-\infty}^\infty dt\Omega(t)$. Note that unlike in the case of $U_Z(\phi)$, here the control fields have to be phase-locked---the expression for $U_{XY}(\theta,\phi)$ depends on the relative phase $\phi$ of the pulses. When $\phi=0$ and $\theta=2\pi$, the evolution operator in Eq.~(\ref{eq:1Q:XY}) performs a {\it single-qubit} swap operation ($X$ gate); i.e., $U=\sigma_x$. The $X$ gate can also be achieved by a resonant pulse sequence of three consecutive $\pi$ pulses (population inversion) involving $\v{e_0}$: $\{\{\Lambda_{R},\pi,0\},\{\Lambda_{B},\pi,0\},\{\Lambda_{R},\pi,0\}\}$ or $\{\{\Lambda_{B},\pi,0\},\{\Lambda_{R},\pi,0\},\{\Lambda_{B},\pi,0\}\}$, where for each pulse, the first element denotes the targeted transition, the second element represent the strength/type of the pulse, and the third is the detuning. We will omit the latter when it is zero for clarity. Both sequences operate in a similar manner by performing three consecutive swaps of population: $\v{0\xi}\leftrightarrow\v{``e"\xi}$, $\v{1\xi}\leftrightarrow\v{``e"\xi}$, and $\v{0\xi}\leftrightarrow\v{``e"\xi}$, which is equivalent to a single $\v{0\xi}\leftrightarrow\v{1\xi}$ swap. Note that in both cases the auxiliary state is assumed to be unpopulated initially and remains empty after the application of the gate.

\section{Two qubit gates}
\label{sec:2Q}

Two-qubit gates rely on physical interactions between the states associated with different qubit systems. In our case the interaction is provided by cavity photons.\cite{Mandel} There are several parameters that characterize the interaction with cavity photons in our two-qubit system. The first, $\Delta$, is the difference in the transition frequencies between the qubit-1 and qubit-2 systems defined in Fig.~\ref{fig:2QC:spec}, i.e., $\Delta = (\varepsilon^{(2)}_{e_0}-\varepsilon^{(2)}_0)-
(\varepsilon^{(1)}_{e_0}-\varepsilon^{(1)}_0)$. The second parameter is the (cavity mode) detuning $\delta = \omega_C - (\varepsilon^{(1)}_{e_0}-\varepsilon^{(1)}_0)$. When $\delta\to   
0$ (or $\delta\to \Delta$) the cavity photon is in resonance with the transition to the first auxiliary state of the qubit-1 (or qubit-2) system. The third parameter is the strength of the coupling between the cavity mode and each qubit system, $g$.

\subsection{off-resonance}
\label{sec:2Q:off}

When $\Delta/g \gg 1$ any interaction between the qubit systems mediated by the cavity is not effective because $\omega_C$ is necessarily far off-resonance with transitions in at least one of the qubit systems. In this regime the qubits are effectively isolated from one another, although each individual qubit system can form superpositions states involving cavity photons when $\delta\to 0$ or $\delta\to\Delta$ for the first or the second qubit system respectively. In this case any pulses applied to the system perform only single-qubit rotations outlined in the previous section.

\subsection{strong resonance}
\label{sec:2Q:in}

When $\Delta/g\ll 1$, the transitions in both qubit systems can be in resonance with the cavity frequency. The spectrum of the system in this case involves entangled combinations of excited states that belong to different qubit systems, e.g., $\v{e_00}\pm\v{0e_0}$. Pulses involving such states will naturally modify entanglement.\cite{economou-2qb} Single-qubit gates, however, are difficult in such systems. A dynamical tuning between strong resonance and off-resonance regimes is used\cite{Lucero} in most systems to perform both single- and two-qubit gates. Dynamical decoupling protocols can also be used in some cases to perform single-qubit operations in always-interacting systems.\cite{Pryadko} Both methods, however, can reduce the efficiency of the gates substantially. 

\subsection{intermediate resonance regime}
\label{sec:2Q:inter}

\begin{figure}
\includegraphics[width=0.9\columnwidth]{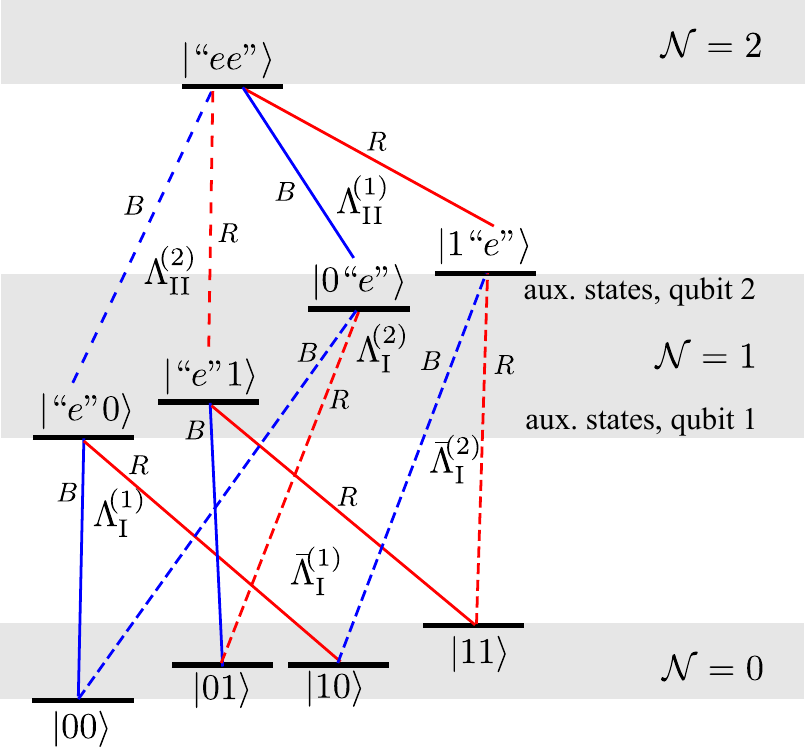}
\caption{\label{fig:2Q:lambdas}
The ladder of transitions for two qubit systems and the cavity grouped into $\Lambda$-systems. Only transitions to one auxiliary state of each qubit system, $\v{e^{(1)}}$ and $\v{e^{(2)}}$, are shown for clarity.
}
\end{figure}

A multi-state qubit-cavity system can have an additional regime that can be used to access both single- and two-qubit functionality efficiently. In order to define this regime we will focus on the first excited state $\v{e_0}\equiv\v{e}$. Transitions to other states have similar structure. In Fig.~\ref{fig:2Q:lambdas} we sketch all transitions involving state $\v{e}$ in $\N=1$ and $\N=2$. These transitions can be grouped into several $\Lambda$-systems. The $\lI{n}$ systems involve states in $\N=0$ and $\N=1$ subspaces and correspond to transitions that are asymptotically connected (at $g\to 0$) to the single-qubit excitations in only one ($n$-th) qubit system. The $\lII{n}$ systems involve states in $\N=1$ and $\N=2$ subspaces and correspond to exciting the $n$-th qubit system when excitation is already present in the other qubit system. The lower frequency leg of each $\Lambda$-system is denoted by R (red) and the higher frequency leg by B (blue). Note that each $\lI{n}$-system has its pair $\lpI{n}$ that is identical to it in the limit $g\to 0$. For example, when $g\to 0$, transition $\LI{1}{B}$ between states $\v{00}\leftrightarrow\v{e0}$ is identical to $\LpI{1}{B}$ between states $\v{01}\leftrightarrow\v{e1}$. When a $\pi$ pulse (population inversion) is applied to these transitions, the initially unentangled qubit state remains unentangled:
\begin{eqnarray}\label{eq:2Q:N1-factr-g0}
&&(A_0\v{0}+A_1\v{1})\otimes(B_0\v{0}+B_1\v{1})
\\\nonumber
&&\xrightarrow[]{\LI{1}{B}} 
A_0B_0i\v{e0}+A_0B_1\v{01}+A_1B_0\v{10}+A_1B_1\v{11}
\\\nonumber
&&\xrightarrow[]{\LpI{1}{B}} 
A_0B_0\v{00}+A_0B_1i\v{e1}+A_1B_0\v{10}+A_1B_1\v{11}
\\\nonumber
&&\xrightarrow[g=0]{\LI{1}{B}\&\LpI{1}{B}} 
(A_0i\v{e}+A_1\v{1})\otimes(B_0\v{0}+B_1\v{1}).
\end{eqnarray}
When $g$ is finite (non-zero), transitions $\LI{1}{B}$ and $\LpI{1}{B}$ are not the same in both the transition energies and the composition of the states (due to mixing with the cavity photon states). In this case, if we tune the $\pi$ pulse to address $\LI{1}{B}$ and perform $\v{00}\to i\v{``e"0}$, the $\LpI{1}{B}$ transition will be off-resonance resulting in incomplete transfer of population and possible accumulation of phase $\v{01}\to\alpha i\v{``e"1}+\beta\v{01}$. The transformation outlined in Eq.~(\ref{eq:2Q:N1-factr-g0}) changes to
\begin{eqnarray}\label{eq:2Q:N1-factr-g1}
&&(A_0\v{0}+A_1\v{1})\otimes(B_0\v{0}+B_1\v{1})
\\\nonumber
&&\xrightarrow[g\ne 0]{\LI{1}{B}\&\LpI{1}{B}}
(A_0\alpha i\v{``e"}+A_1\v{1})\otimes(B_0\v{0}+B_1\v{1})
\\\nonumber
&&+A_0B_0(1-i\alpha)\v{00}+A_0B_1\beta\v{01}.
\end{eqnarray}
This state is not factorizable and is entangled.\cite{nielsenchuang,Wootters} The same transformation can be performed using the $R$ transitions, as well as the transitions in $\lI{2}$ (to address the second qubit). The degree of the entanglement generation can be judged by the detuning $\Delta\omega_{\N=1}$ of transition $\LI{1}{B}$ from $\LpI{1}{B}$ formulated in Eq.~(\ref{eq:2QC:omega-N1}). The value of $\Delta\omega_{\N=1}$ decreases as the spectral distance between transitions in the qubit-1 and qubit-2 systems, $\Delta$, increases (see Fig.~\ref{fig:2QC:int} and Appendix~\ref{app:interaction}). This stems from the fact that a cavity photon emitted from one of the qubit systems is detuned from the transition in the other and cannot be effectively reabsorbed, resulting in attenuation of the exchange interaction between the two systems. When $\delta \sim \Delta$, the transition energy difference $\Delta\omega_{\N=1}$ can approach $\sim g^{n+1}/\Delta^n$ with $n>2$ (see Fig~\ref{fig:2QC:int}, Appendix~\ref{app:interaction}, and Sec.~\ref{sec:SC} for details).

For the same reason, when $g\to 0$, transitions in $\lII{n}$ are identical to those in $\lI{n}$ and cannot create entanglement. Consider, for instance, changes to an initially unentangled state due to transitions $\LII{2}{B}$ and $\LpI{2}{B}$:
\begin{eqnarray}\label{eq:2Q:N2-factr-g0}
&&(A_0\v{e}+A_1\v{1})\otimes(B_0\v{0}+B_1\v{1})
\\\nonumber
&&\xrightarrow[]{\LII{2}{B}} 
A_0B_0i\v{ee}+A_0B_1\v{e1}+A_1B_0\v{10}+A_1B_1\v{11}
\\\nonumber
&&\xrightarrow[]{\LpI{2}{B}} 
A_0B_0\v{e0}+A_0B_1\v{e1}+A_1B_0i\v{1e}+A_1B_1\v{11}
\\\nonumber
&&\xrightarrow[g=0]{\LII{2}{B}\&\LpI{2}{B}} 
(A_0\v{e}+A_1\v{1})\otimes(B_0i\v{e}+B_1\v{1}).
\end{eqnarray}
When $g$ is finite (non-zero) and we tune the $\pi$ pulse in resonance with $\LII{2}{B}$, instead of Eq.~(\ref{eq:2Q:N2-factr-g0}) we obtain the transformation
\begin{eqnarray}\label{eq:2Q:N2-factr-g1}
&&(A_0\v{``e"}+A_1\v{1})\otimes(B_0\v{0}+B_1\v{1})
\\\nonumber
&&\xrightarrow[g=0]{\LII{2}{B}\&\LpI{2}{B}} 
(A_0\v{``e"}+A_1\v{1})\otimes(B_0\alpha' i\v{``e"}+B_1\v{1})
\\\nonumber
&&+A_0B_0(1-i\alpha')\v{``e"0}+A_0B_1\beta'\v{10},
\end{eqnarray}
in which entanglement is generated. Note that here the pulses are applied to the second qubit (qubit-2 system), and, due to differences in $\LII{2}{B}$ and $\LpI{2}{B}$, the transformation for state $\v{10}$ is incomplete $\v{10}\to\alpha i\v{1``e"}+\beta\v{10}$.

The transformations (\ref{eq:2Q:N1-factr-g1}) and (\ref{eq:2Q:N2-factr-g1}) are similar except that the latter addresses the second qubit and the first qubit has already been excited. However, there is a crucial difference in the degree of entanglement that these two transformations generate for different $g$, $\Delta$, and $\delta$. To compare them we construct the transition energy difference $\Delta\omega_{\N=2}$ based on transformation (\ref{eq:2Q:N2-factr-g1}), or the second and the third lines of (\ref{eq:2Q:N2-factr-g0}) [see also Eq.~(\ref{eq:2QC:omega-N2})], in a fashion similar to $\Delta\omega_{\N=1}$. Both $\Delta\omega_{\N=1}$ and $\Delta\omega_{\N=2}$ are zero when $g/\Delta\to 0$ for the reasons mentioned above. When $g/\Delta$ is non-zero, $\Delta\omega_{\N=1}$ and $\Delta\omega_{\N=2}$ can deviate from one another substantially (see Fig.~\ref{fig:2QC:int}). At $\delta=\Delta$ we obtain $\Delta\omega_{\N=2}\sim g^2/2\Delta$, and, thus, $\Delta\omega_{\N=1}/\Delta\omega_{\N=2}\sim (g/\Delta)^{n-1}$ with $n> 2$, as shown analytically in Appendix~\ref{app:interaction} and Sec.~\ref{sec:SC}. The results of the numerical analysis presented in Fig.~\ref{fig:2QC:int} show that this trend remains valid up to $g/\Delta$ of about $1/4$. Therefore at some intermediate values of $g/\Delta$ all transformations (gates) that do not involve $\N=2$ states are local with good accuracy, whereas transformations (gates) that reach at least one of the $\N=2$ states can be non-local and can generate entanglement.

\subsubsection{non-commutativity of single-qubit pulses}
\label{sec:2Q:non-com}

An interesting consequence of a substantial difference between $\Delta\omega_{\N=1}$ and $\Delta\omega_{\N=2}$ in the intermediate regime of resonance is non-commutativity of single-qubit manipulations. Consider, for example, two simple diagonal single-qubit operations (Z gates) performed via excited state $\v{``e"}$ and applied to the first and, then, the second qubit. As was shown in Sec.~\ref{sec:1Q}, each of these Z gates can be executed by two $\pi$ pulses (or a single combined $2\pi$ pulse), i.e., $\{\{\LI{1}{B},\pi\},\{\LI{1}{B},\pi\}\}=\{\LI{1}{B},2\pi\}$ for the first qubit and $\{\{\LI{2}{B},\pi\},\{\LI{2}{B},\pi\}\}=\{\LI{2}{B},2\pi\}$ for the second qubit. In each case, the first $\pi$ pulse brings the population of the state $\v{0}$ to $\v{``e"}$, and the second pulse returns it to $\v{0}$ leaving $\v{``e"}$ empty, as it was initially. Note, however, that if we attempt to perform a Z gate on the second qubit in between the two pulses that produce the Z gate for the first qubit the result will be different; i.e.,
\begin{eqnarray}\label{eq:2Q:non-com-example-a}
\{\{\LI{1}{B},\pi\};\{\LI{2}{B},\pi\},\{\LI{2}{B},\pi\};\{\LI{1}{B},\pi\}\}
\\\label{eq:2Q:non-com-example-b}
\neq
\{\{\LI{1}{B},2\pi\},\{\LI{2}{B},2\pi\}\}.
\end{eqnarray}
The action of the first pulse in Eq.~(\ref{eq:2Q:non-com-example-a}) is outlined in Eq.~(\ref{eq:2Q:N1-factr-g0}). The second and the third pulses operate on the second qubit. They start with the state in which $\v{00}\to\v{``e"0}$, and, thus, their operation is given by Eq.~(\ref{eq:2Q:N2-factr-g1}) instead of Eq.~(\ref{eq:2Q:N1-factr-g0}). Because $\Delta\omega_{\N=1}$ is negligible and $\Delta\omega_{\N=2}$ is not, the second and the third pulses start generating entanglement and are no longer a single qubit manipulation as they are in Eq.~(\ref{eq:2Q:non-com-example-b}).

More generally it can be formulated as follows: {\it any local (single-qubit) operation that uses auxiliary states cannot, in general, remain local if it is interchanged or coincides in time with another local operation that leaves population in at least one of the same auxiliary states}.

\begin{figure}
\includegraphics[width=0.8\columnwidth]{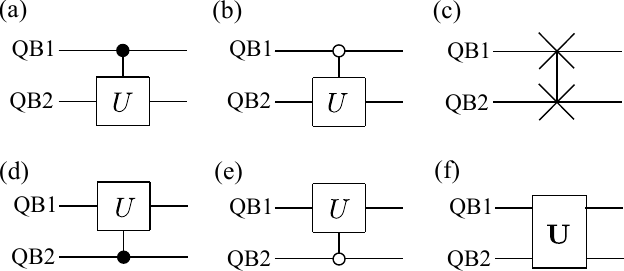}
\caption{\label{fig:2Q:CU}
Two-qubit gates. Panels (a) and (b) show $C(U)$ (control-U) gates in which $U$ is applied to the second qubit if the first qubit is in state $\v{1}$ and $\v{0}$ respectively. Panel (c) shows the {\it two-qubit} swap gate (SWAP), i.e., the populations of the first qubit (QB1) and the second qubit (QB2) are interchanged.  Panels (d) and (e) show $C(U)$ gates in which $U$ is applied to the first qubit if the second qubit is in state $\v{1}$ and $\v{0}$ respectively. Finally, panel (f) represents a general two-qubit rotation. 
}
\end{figure}

This non-commutativity of single-qubit controls can be the basis for a variety of two-qubit operations shown in Fig.~\ref{fig:2Q:CU}. Each gate in Fig.~\ref{fig:2Q:CU} can be performed by a specifically ordered sequence of pulses, each corresponding to a simple local (single-qubit) operation in one of the qubit systems in the intermediate resonance regime. In what follows, we will focus only on the intermediate regime of resonance in which $g/\Delta$ is sufficiently small so that $\Delta\omega_{\N=1}$ is negligible and yet sufficiently large so that $\Delta\omega_{\N=2}$ can lead to substantial phase accumulation for the desired time interval.

\subsubsection{gates}
\label{sec:2Q:gates}

One of the most important classes of two-qubit operations that is a consequence of the non-commutativity of single-qubit controls is $C(U)$ (control-U) operations,\cite{nielsenchuang} such as
\begin{equation}\label{eq:2Q:CU}
C(U) = \left({\small
\begin{array}{cccc}
\begin{array}{cccc}
1 & 0\\
0 & 1\\
\end{array}
& 
\begin{array}{cccc}
0 & 0\\
0 & 0\\
\end{array}
\\
\begin{array}{cccc}
0 & 0\\
0 & 0\\
\end{array}
&
\mathlarger{\mathlarger{\mathlarger{\mathlarger{U}}}}
\\
\end{array}
}\right),
\end{equation}
where $U$ is an arbitrary $2\times 2$ unitary matrix. Using the $n$-th qubit as a control qubit and the $m$-th qubit as a target qubit, $C(U)$ is performed by a series of pulses 
\begin{eqnarray}\label{eq:2Q:CU-pulses-ALL}
\{\{\LI{n}{i},2\pi M+\pi\}, [\Lambda^{(m)}_{\rm I\,\,or\,\,II}],\{\LI{n}{i},2\pi M'+\pi\}\}, 
\end{eqnarray}
where $[\Lambda^{(m)}_{\rm I\,\,or\,\,II}]$ is a single two-color pulse or a series of simple pulses that would normally perform a single-qubit operation $U$ on the $m$-th qubit using the corresponding $\Lambda$-systems (see Sec.~\ref{sec:1Q}). Here $i$ is used to refer to $B$ or $R$. The first and the last pulses are resonant $\pi$ (swap or population inversion) pulses, with additional $2\pi M$ and $2\pi M'$ rotations that are used to correct for incurred single-qubit phases. The sequence (\ref{eq:2Q:CU-pulses-ALL}) supports several variations
\begin{eqnarray}\label{eq:2Q:CU-pulses-eg1}
\{\{\LI{1}{B},2\pi M+\pi\}, [\lI{2}],\{\LI{1}{B},2\pi M'+\pi\}\}, 
\\\label{eq:2Q:CU-pulses-eg2}
\{\{\LI{1}{R},2\pi M+\pi\}, [\lII{2}],\{\LI{1}{R},2\pi M'+\pi\}\}, 
\\\label{eq:2Q:CU-pulses-eg3}
\{\{\LI{1}{B},2\pi M+\pi\}, [\lII{2}],\{\LI{1}{B},2\pi M'+\pi\}\}.
\end{eqnarray}
The first two sets of pulses, (\ref{eq:2Q:CU-pulses-eg1}) and (\ref{eq:2Q:CU-pulses-eg2}), will apply $U$ to the second qubit if the first (control) qubit is in state $\v{1}$; see Eq.~(\ref{eq:2Q:CU}) and Fig.~\ref{fig:2Q:CU}(a). The third set [Eq.~\ref{eq:2Q:CU-pulses-eg3}] will apply $U$ to the second qubit only if the control qubit is in state $\v{0}$; see Fig.~\ref{fig:2Q:CU}(b). Control gates with the second qubit as a control qubit [Figs.~\ref{fig:2Q:CU}(d) and (e)] can be obtained by replacing $\Lambda^{(1)} \leftrightarrow \Lambda^{(2)}$ in the pulse sequences.

A hierarchy of $\Lambda$ systems in the intermediate resonance regime can also be used to perform other two-qubit gates that are not $C(U)$ gates. For example, a {\it two-qubit} swap operation,\cite{nielsenchuang} 
\begin{eqnarray}\label{eq:2Q:SWAP-pulses}
{\rm SWAP} = 
\left(
\begin{array}{cccc}
1 & 0 & 0 & 0\\
0 & 0 & 1 & 0\\
0 & 1 & 0 & 0\\
0 & 0 & 0 & 1\\
\end{array}
\right),
\end{eqnarray}
can be performed by a sequence
\begin{eqnarray}\label{eq:2Q:SWAP-pulses}
\{
\{\LI{1}{B},\pi\}, 
\{\LI{2}{B},\pi\}, 
[
\{\LII{2}{R},\pi\}, 
\{\LII{1}{R},\pi\},
\\\nonumber
\{\LII{2}{R},3\pi\}
],
\{\LI{2}{B},\pi\}, 
\{\LI{1}{B},3\pi\}
\}.
\end{eqnarray}
Note, that the phase difference, $\phi$, for the three middle pulses originating from different phases of the pulse fields will enter the result, as happens for the {\it single-qubit} swap (X) gate (see Sec.~\ref{sec:1Q} and Appendix~\ref{app:phase-lock}). Therefore, pulses addressing $\LII{2}{R}$ and $\LII{1}{R}$ transitions have to be phase-locked (zero phase difference, $\phi=0$).

In order to understand the composition of the pulse sequences, we examine sequence (\ref{eq:2Q:CU-pulses-eg1}) in greater detail. We begin with an arbitrary two-qubit state
\begin{eqnarray}\nonumber
C_{00}\v{00}+C_{01}\v{01}+C_{10}\v{10}+C_{11}\v{11}. 
\end{eqnarray}
The first resonant pulse $\{\LI{1}{B},2\pi M+\pi\}$ is tuned to transition $B$ of the $\lI{1}$ system. In the intermediate resonance regime the $\N=1$ states remain local to each qubit. As a result, transition $B$ of $\lI{1}$ is indistinguishable from the $B$ of $\lpI{1}$, and the populations of state $\v{00}$ and state $\v{01}$ are transferred to $\v{``e"0}$ and $\v{``e"1}$ respectively [see Eq.~(\ref{eq:2Q:N1-factr-g0})]. After this operation only states $\v{10}$ and $\v{11}$ of the qubit sub-space are occupied, and we have
\begin{eqnarray}\nonumber
iC_{00}\v{``e"0}+iC_{01}\v{``e"1}+C_{10}\v{10}+C_{11}\v{11}.\end{eqnarray}
Because the $\N=1$ states are local, $\lI{2}$ and $\lpI{2}$ are also indistinguishable. However, since our state does not have states $\v{00}$ and $\v{01}$ anymore, we can operate only in $\lpI{2}$ and not in $\lI{2}$. The former is a familiar single-qubit $\Lambda$-system and, thus, any single-qubit $U$ can be performed on states $\v{10}$ and $\v{11}$ as if operating on the second qubit alone, i.e. $U[C_{10}\v{10}+C_{11}\v{11}]$. In other words, the pulses that perform $U$ are exactly the same pulses that one would use to apply $U$ to the second qubit without any prior manipulations with the first qubit. The states $\v{``e"0}$ and $\v{``e"1}$ are not affected by near-resonant pulses operating in $\lpI{2}$ because transitions in $\lII{2}$ are sufficiently detuned from $\lpI{2}$ due to the interaction with the cavity photons. Therefore, after pulses $[\lI{2}]$ are applied, we have
\begin{eqnarray}\nonumber
iC_{00}\v{``e"0}+iC_{01}\v{``e"1}+U[C_{10}\v{10}+C_{11}\v{11}]. \end{eqnarray}
Finally, the last resonant pulse (identical to the first one) transfers the population back to the qubit subspace, and we obtain 
\begin{eqnarray}\nonumber
-C_{00}\v{00}-C_{01}\v{01}+U[C_{10}\v{10}+C_{11}\v{11}], \end{eqnarray}
which concludes the operation of the gate. The minus signs acquired before states $\v{00}$ and $\v{01}$ can be removed by adding a $2\pi M$ ($M$ is a non-zero integer) to the first or the last pulse if necessary. Note that similar $C(U)$ operation can be performed using $\lII{2}$, Eq.~(\ref{eq:2Q:CU-pulses-eg2}), instead of $\lpI{2}$. The latter $\Lambda$ system, however, is likely to be more convenient experimentally: the same $\lpI{2}$ is used in single-qubit manipulations; thus no additional calibration is needed, and single-qubit pulses used for individual single-qubit gates can be used for $C(U)$.

Pulses operating in all $\Lambda$-systems can be combined into a single multi-color pulse with a complex shape. Such a pulse can perform an arbitrary two-qubit operation [Fig.~\ref{fig:2Q:CU}(f)]. The shape of the multi-color pulse can potentially be better optimized for each operation, thus reducing the total gate time relative to that of the simple sequences of resonant pulses discussed above.

The gates proposed here can be performed with the desired accuracy in a fully coherent system by choosing sufficiently small bandwidths for the pulses. It is evident, however, that in a realistic environment the duration of pulses is limited  by decoherence.\cite{nielsenchuang,vanKampen,Leggett} As a result, the value of $\Delta\omega_{\N=2}$ has to be sufficiently large, and the strong coupling regime with respect to the decoherence rates,\cite{solenov1,solenov2} i.e., $\Delta\omega_{\N=2}\gg \max[\Gamma_i]$, is necessary.
In the following sections we will demonstrate that the proposed two-qubit gates can be performed in different physical systems with fidelity in excess of 90\% for $g/\max[\Gamma_i]\gtrsim 100$ and $g/\Delta\sim 0.1$.

\section{Self-assembled quantum dots}
\label{sec:QD}

\begin{figure}
\includegraphics[width=0.99\columnwidth]{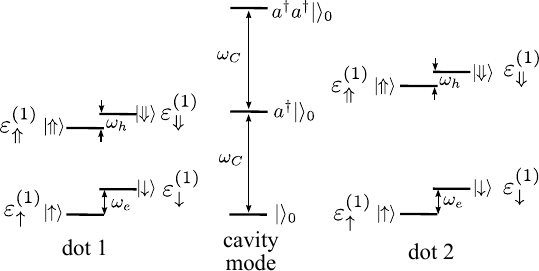}
\caption{\label{fig:QD:Levels}
Relevant energy levels for two non-identical charged quantum dots and a photonic microcavity (only the three lowest states of the harmonic ladder are shown). Negatively charged exciton (trion) states, $\v{\U}$ and $\v{\D}$, are composed of two electrons and one hole and have quasi-spin $1/2$ (here we neglect mixing between heavy and light holes). The states $\v{\u}$ and $\v{\d}$ are spin states of the electron in each quantum dot. A magnetic field perpendicular to the growth direction is assumed (spins are quantized in the direction of the magnetic field). The magnetic g-factor is assumed to be the same for electrons in both quantum dots; however, $\omega_h\neq\omega_e$ ($\delta\omega=\omega_h-\omega_e\sim \omega_e$).
}
\end{figure}

In this section we review an example of the system of two negatively charged quantum dots interacting with a microcavity mode investigated earlier in Ref.~\onlinecite{solenov-QD}. This example focuses on a specific CZ gate, ${\rm diag}\{1,-1,1,1\}$, to illustrate the more general formulation of arbitrary two-qubit operations developed in the previous section.

In a system of a charged self-assembled InAs/GaAs quantum dot the qubit is typically encoded by the spin of an extra electron.\cite{Greilich} Auxiliary states $\v{e_i}$ are due to charged exciton states (trions) that include two electrons and one heavy hole. Mixing between heavy and light hole spin states is neglected. In this case, the trion has two spin (quasispin) states approximately corresponding to those of the heavy hole. We will use spin notation for qubit and auxiliary states by setting $0=\u$, $1=\d$, $e_0=\U$, and $e_1=\D$ (see Fig. \ref{fig:QD:Levels}). The transition energy $\omega_{\u\leftrightarrow\U}$ is controlled by the size of each dot and varies naturally with the typical width of the distribution $\sim 10$meV observed during the growth. In experiment, self-assembled dots are post-selected and the spectral variations for the two-dot system can be reduced to $\Delta\sim 1$meV. Trion transitions in each quantum dot are spin-conserving with respect to the spin projection onto the growth axis direction. They couple only to optical photons with in-plane polarization due to substantial crystal strain in the InAs dots. 

To provide full control over the spin states a $\Lambda$ system is formed by applying a magnetic field in-plane. The spin states projected onto the external magnetic field contain both projections on to the growth direction and, thus, $\u\leftrightarrow\D$ and $\d\leftrightarrow\U$ transitions become possible. Transitions $\v{\u}\leftrightarrow\v{\U}$ and $\v{\d}\leftrightarrow\v{\D}$ are coupled to photons with vertical (along the magnetic field) polarization, and transitions $\v{\u}\leftrightarrow\v{\D}$ and $\v{\d}\leftrightarrow\v{\U}$ are coupled to photons with horizontal (perpendicular to the magnetic field) polarization.

The cavity is typically formed as a defect in a photonic crystal membrane.\cite{Carter} We will consider the case, in which only $\v{\u}\leftrightarrow\v{\U}$ and $\v{\d}\leftrightarrow\v{\D}$
are coupled to the cavity mode for clarity. We also assume that the coupling constant $g=0.4$meV is the same for both qubits. In this case, the potential for generating entanglement via $\N=1$ states is negligible: the value of $\Delta\omega_{\N=1}$, which is at most $\sim\omega_e g^2/\Delta^2$ (see Appendix~\ref{app:interaction}), is well below the spontaneous recombination rate of trion transitions, $\Gamma_{\rm trion}\sim 0.5 \mu$eV. At the same time $\Delta\omega_{\N=2}$ remains sufficiently large to perform entangling gates in the intermediate regime of resonance.

\begin{figure}
\includegraphics[width=0.9\columnwidth]{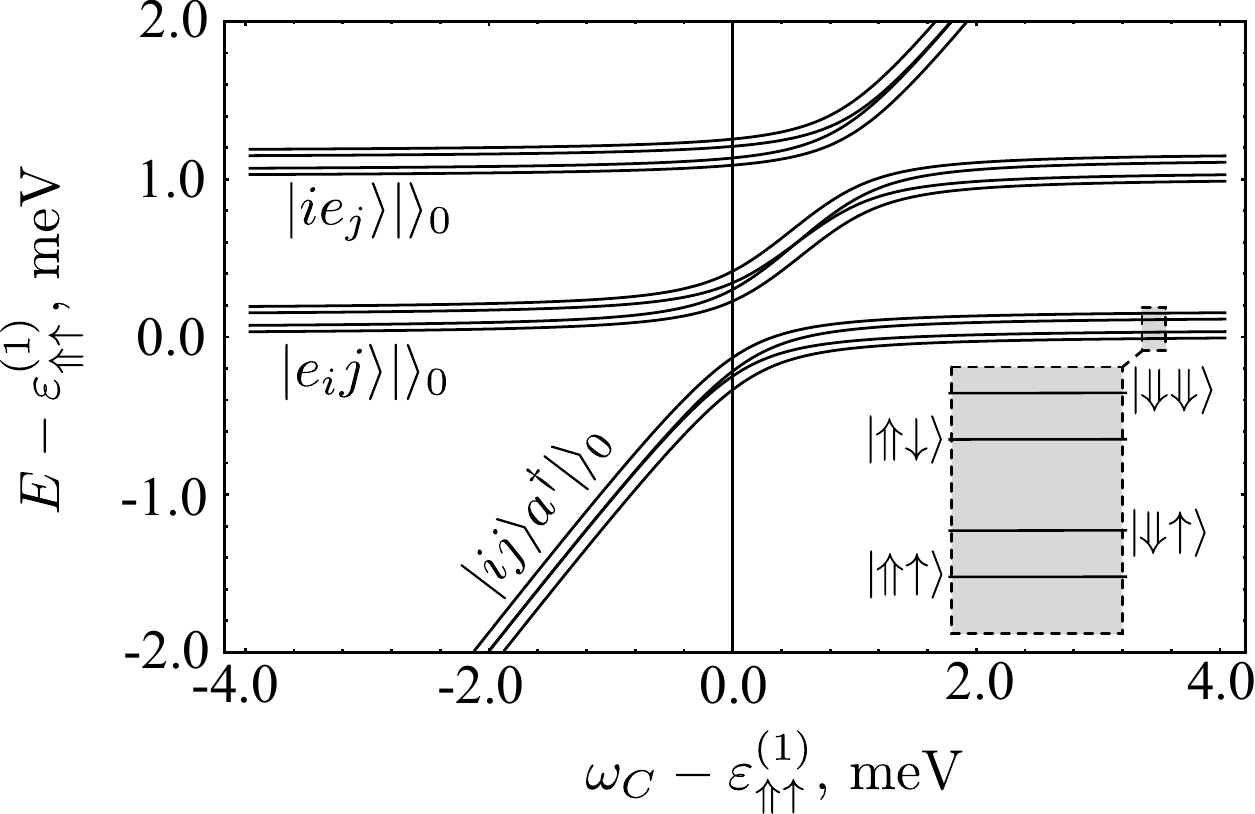}
\caption{\label{fig:QD:N1}
One excitation sector ($\N=1$) of the spectrum in the system of two negatively charged quantum dots interacting with microcavity photons (obtained numerically). Spin (quasi-spin) notations are used, $0=\u$, $1=\d$, $e_0=\U$, $e_1=\D$. In the plot, $\omega_e=3\omega_h=0.12$meV, $g=0.4$meV, $\Delta=1$meV}
\end{figure}
\begin{figure}
\includegraphics[width=0.9\columnwidth]{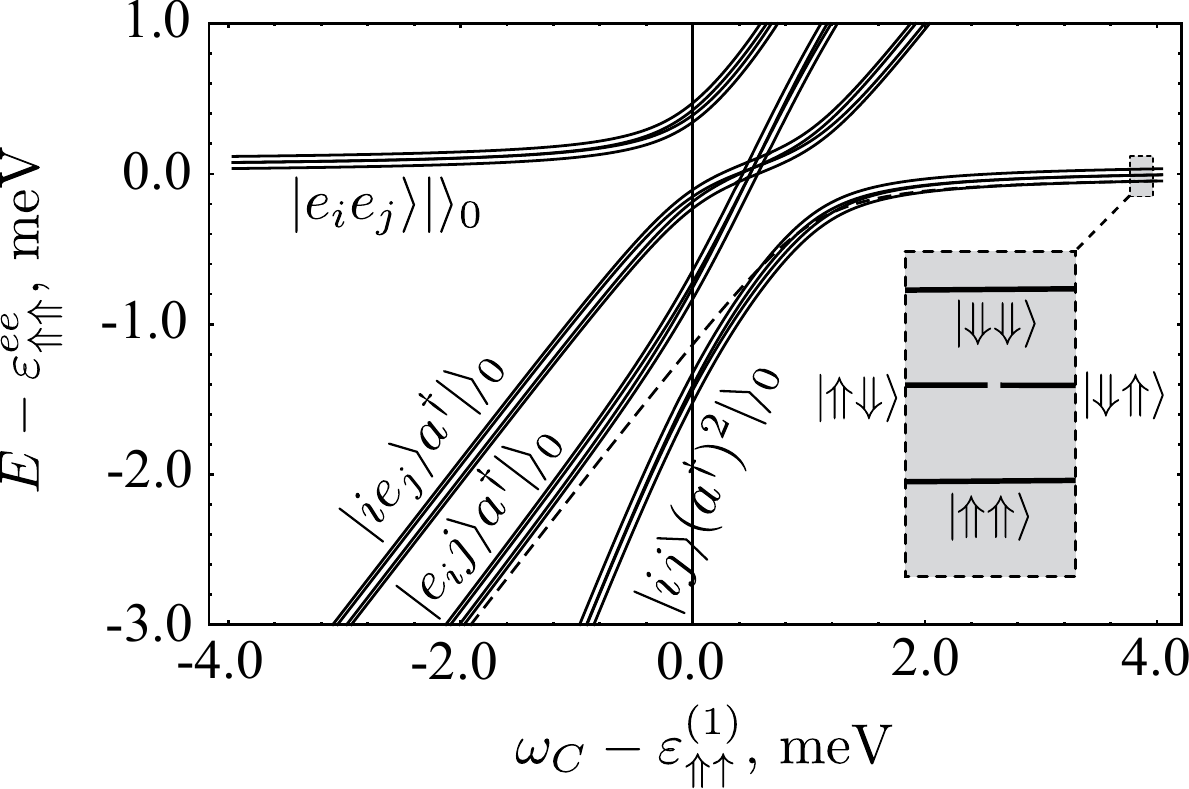}
\caption{\label{fig:QD:N2}
Two-excitation sector ($\N=2$) of the spectrum in the system of two negatively charged quantum dots interacting with microcavity photons (obtained numerically). Parameters are the same as in Fig.~\ref{fig:QD:N1}.
}
\end{figure}
\begin{figure}
\includegraphics[width=0.9\columnwidth]{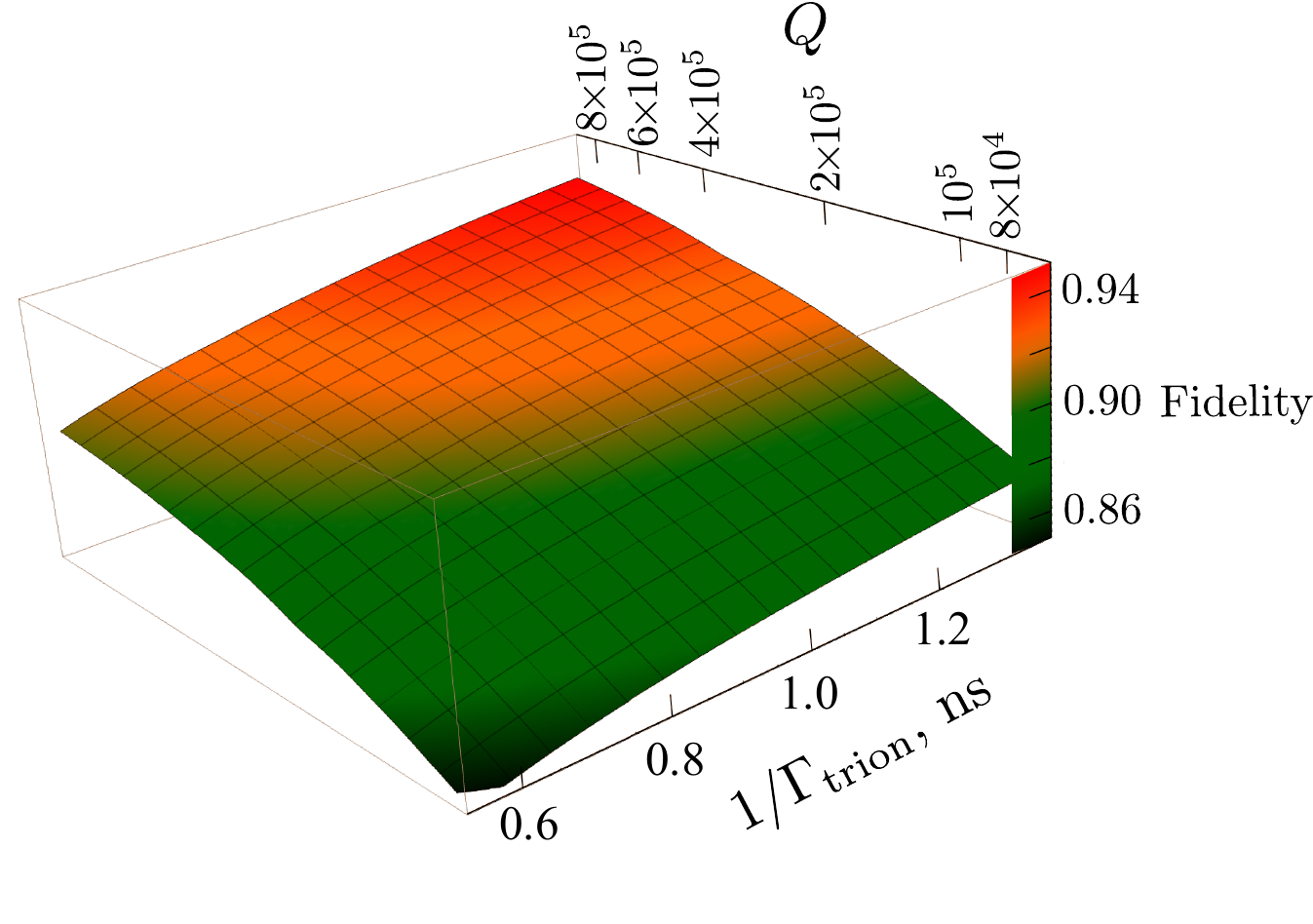}
\caption{\label{fig:QD:fidel}
Calculated fidelity of the two-qubit CZ gate in the system of two negatively charged quantum dots interacting with a microcavity mode. Parameters are the same as in Fig.~\ref{fig:QD:N1}.
The area with fidelities exceeding 90\% is highlighted in red. Each point of the surface reflects fidelity optimized over $\delta$, and with respect to the bandwidth of the pulses (see the text and Ref.~\onlinecite{solenov-QD} for details).
}
\end{figure}

The spectrum of the system of two quantum dots and a cavity mode is computed\cite{solenov-QD} numerically and presented in Figs. \ref{fig:QD:N1} and \ref{fig:QD:N2}. The first band in the $\N=1$ sector (see Fig. \ref{fig:QD:N1}) has four states with $\U\u$, $\D\u$, $\U\d$, $\D\d$ spin configurations (bottom to top). When $\omega_C$ is close to the trion transition energy, these states form local polaron-like complexes with the cavity photon states $\v{\u\u}a^\dag\v{}_0$, $\v{\u\d}a^\dag\v{}_0$, $\v{\d\u}a^\dag\v{}_0$, and $\v{\d\d}a^\dag\v{}_0$ (diagonal lines in Fig. \ref{fig:QD:N1}). A similar set of anti-crossings is formed for the second band. In the $\N=2$ sector, shown in Fig. \ref{fig:QD:N2}, two-trion states ($\v{\U\U}$, $\v{\U\D}$, $\v{\D\U}$ and $\v{\D\D}$) are coupled to states with one and two cavity photons.

We calculate\cite{solenov-QD} the fidelity for the $CZ={\rm diag}\{1,-1,1,1\}$ gate outlined in Eq.~(\ref{eq:2Q:CU-pulses-eg3}). The gate is based on transitions $\LI{1}{B}$ and $\LII{2}{B}$, and we take $M=M'=0$ (see Secs.~\ref{sec:1Q} and \ref{sec:2Q:gates}). The single middle pulse in Eq.~(\ref{eq:2Q:CU-pulses-eg3}), $\{\LII{2}{B},2\pi\}$, performs the $U=-Z$ operation reaching to $\N=2$ subspace. The transitions $\LI{1}{B}=\LpI{1}{B}$ connect states $\v{\u\u}$ and $\v{\u\d}$ from the $\N=0$ subspace with states $\v{``\U"\u}$ and $\v{``\U"\d}$ from the $\N=1$ subspace, respectively. The latter correspond to the first and the third energy curves in Fig.~\ref{fig:QD:N1} (from the bottom). The transition $\LII{2}{B}$ involves states $\v{``\U"\u}$ and $\v{``\U\U"}$ (the bottom energy curve in Fig.~\ref{fig:QD:N2}). The gate is expected to perform well for $\omega_C\gtrsim\varepsilon^{(1)}_{\U\u}$. In the case of $\omega_C\lesssim\varepsilon^{(1)}_{\U\u}$ similar states to the left of the anti-crossings should be chosen to avoid exciting the cavity mode directly.

The fidelity is averaged (see Appendix~\ref{app:fidel}) with respect to an arbitrary initial two-qubit wave function to provide an estimate suitable for different quantum algorithms. The averaged fidelity as a function of the cavity quality factor $Q$ and trion recombination time $\Gamma_{\rm trion}$ is shown in Fig. \ref{fig:QD:fidel}. Each point in this plot is optimized with respect to the mode frequency $\omega_C$ over the range shown in Figs. \ref{fig:QD:N1} and \ref{fig:QD:N2}, and also over the pulse bandwidth, $\sigma$. The fidelity is typically maximized at values of $\delta=\omega_C-\varepsilon^{(1)}_{\U\u}\sim\Delta$ (see Ref.~\onlinecite{solenov-QD} and Appendix~\ref{app:interaction}).

\section{Defects in diamond and silicon carbide}
\label{sec:NV}

In this section we present a different example in which qubits are encoded by the electronic states of a NV defect in diamond or similar defects in silicon carbide.\cite{solenov-NV} Diamond and silicon carbide have a variety of defect centers that can be accessed optically.\cite{Kennedy,Awschalom-SiC,Acosta,Doherty,Baranova,Gali,Riedel,MazeLukin} We will focus on negatively charged NV centers in diamond that are currently used to demonstrate single-qubit manipulation at room temperature.\cite{Kennedy}

\begin{figure}
\includegraphics[width=0.7\columnwidth]{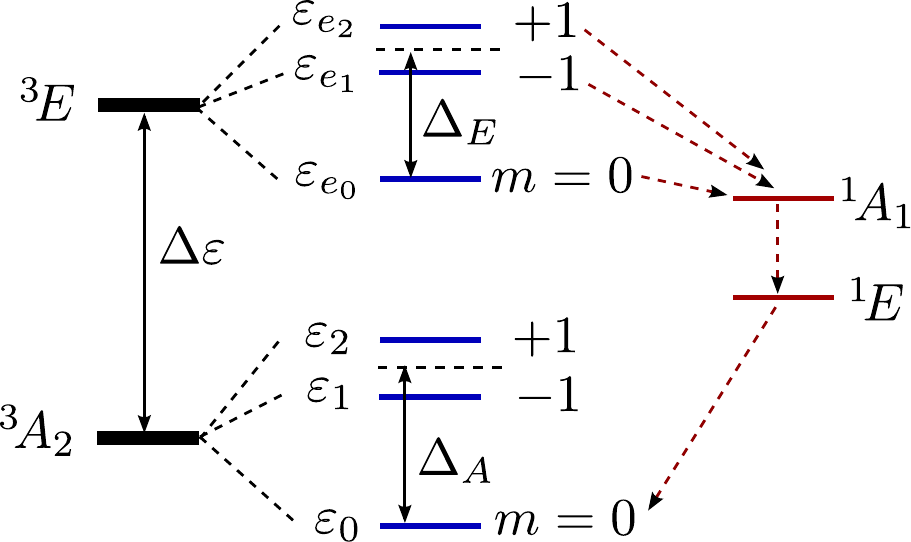}
\caption{\label{fig:NV:levels}
Relevant energy levels of a single NV center in diamond. The qubit is typically encoded by the lowest two states of the $^3A_2$ triplet. The two states $^1A_1$ and $^1E$ participate in non-radiative decay of the upper triplet states.}
\end{figure}
\begin{figure}
\includegraphics[width=0.8\columnwidth]{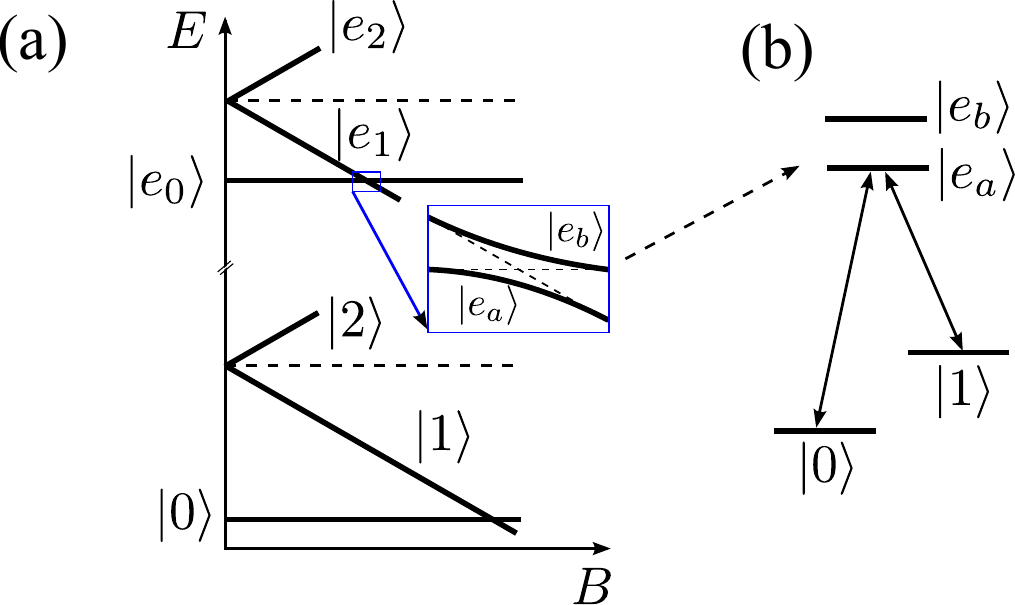}
\caption{\label{fig:NV:cross-levels}
Energies of a singe NV center in magnetic field. (a) Energies as a function of external magnetic field. (b) A $\Lambda$ system formed due to mixing of $m=0$ and $m=-1$ states of the $^3E$ triplet at some finite (non-zero) magnetic field.}
\end{figure}

The relevant states of a NV center\cite{Acosta} are shown in Fig. \ref{fig:NV:levels}. Each triplet is a spin-1 system. The $m=\pm 1$ spin projections are typically split due to local strain in the defect,\cite{Bassett} and the splitting energies $\Delta_A$ and $\Delta_E$ can potentially be tuned further with an external electric field. The first two states of the $^3A_2$ triplet are used to encode the qubit.\cite{Kennedy,Acosta,Doherty,MazeLukin,Togan} Optically accessible $^3E$ triplet states are used to manipulate the qubit. Unlike the case of quantum dots, these excited states have an additional non-radiative decay pathway\cite{Acosta} through $^1A_1$ and $^1E$ with different rates for $m=0$ and $m=-1$ states. These pathways are used to initialize the system to its ground state ($^3A_{2,m=0}$) and to readout the qubit.\cite{Yale} The latter is possible due the difference in the non-radiative decay rate for states $^3E_{m\neq 0}$ and $^3E_{m=0}$. 

\begin{figure}
\includegraphics[width=0.7\columnwidth]{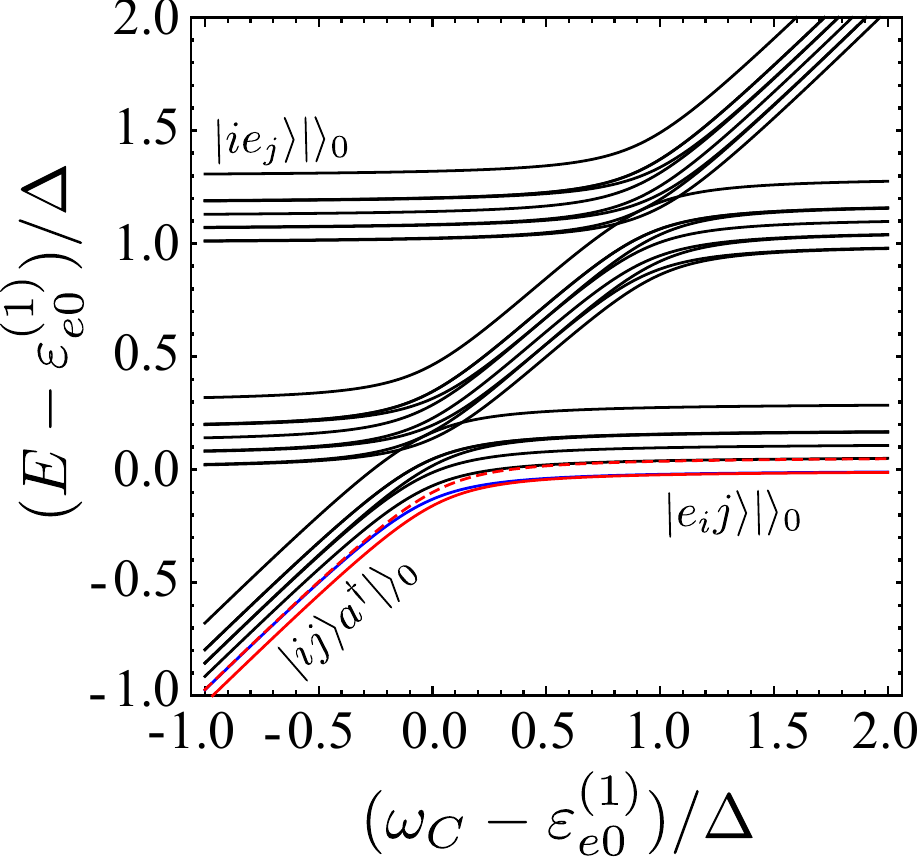}
\caption{\label{fig:NV:N1}
One excitation sector ($\N=1$) of the spectrum in the system of two NV centers interacting with microcavity photons (obtained numerically). We use a shortened notation $\v{e_a}\to \v{e}$, and we use the typical numbers for $\Delta_A$ and $\Delta_E$ (see Ref.~\onlinecite{solenov-NV}). The other parameters are $\Delta=0.1$meV, $g=15\mu$eV, $g_0=0.1\mu$eV.
}
\end{figure}

\begin{figure}
\includegraphics[width=0.8\columnwidth]{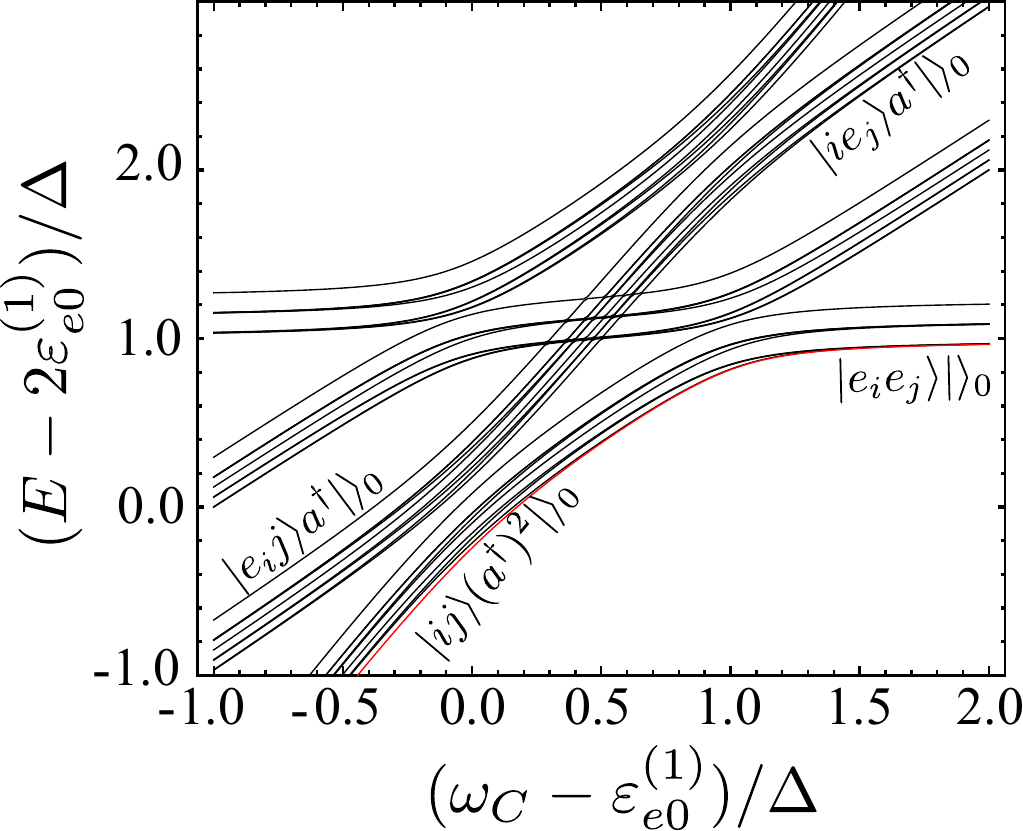}
\caption{\label{fig:NV:N2}
Two-excitation sector ($\N=2$) of the spectrum in the system of two NV centers interacting with microcavity photons (obtained numerically). Parameters are the same as in Fig.~\ref{fig:NV:N1}.
}
\end{figure}

The optical transitions in this system are generally spin conserving ($\v{i}\leftrightarrow\v{e_i}$). However at finite (non-zero) magnetic field, $B$, an anti-crossing between $^3E_{m=0}$ and $^3E_{m=-1}$ can be reached,\cite{Yale,Togan} mixing these two states
\begin{eqnarray}\nonumber
g_0 (\v{e_0}\iv{e_1}+\v{e_1}\iv{e_0}),
\end{eqnarray}
as shown in Fig. \ref{fig:NV:cross-levels}(a). The resulting superposition states are accessible from both $^3A_{2, m=0}$ and $^3A_{2, m=-1}$ qubit states, and a set of $\Lambda$ systems can be formed, as in Fig. \ref{fig:NV:cross-levels}(b).
We will focus on such a case here. To shorten notations we will use index $e$ for the lowest superposition state, $\v{e_a}$.
In Figs. \ref{fig:NV:N1} and \ref{fig:NV:N2} we show the spectrum of two NV-centers interacting with a single cavity mode. The overall structure of the $\N=1$ and $\N=2$ sectors of the spectrum is similar to that in Figs. \ref{fig:QD:N1} and \ref{fig:QD:N2} for InAs/GaAs quantum dots.


\begin{figure}
\includegraphics[width=0.8\columnwidth]{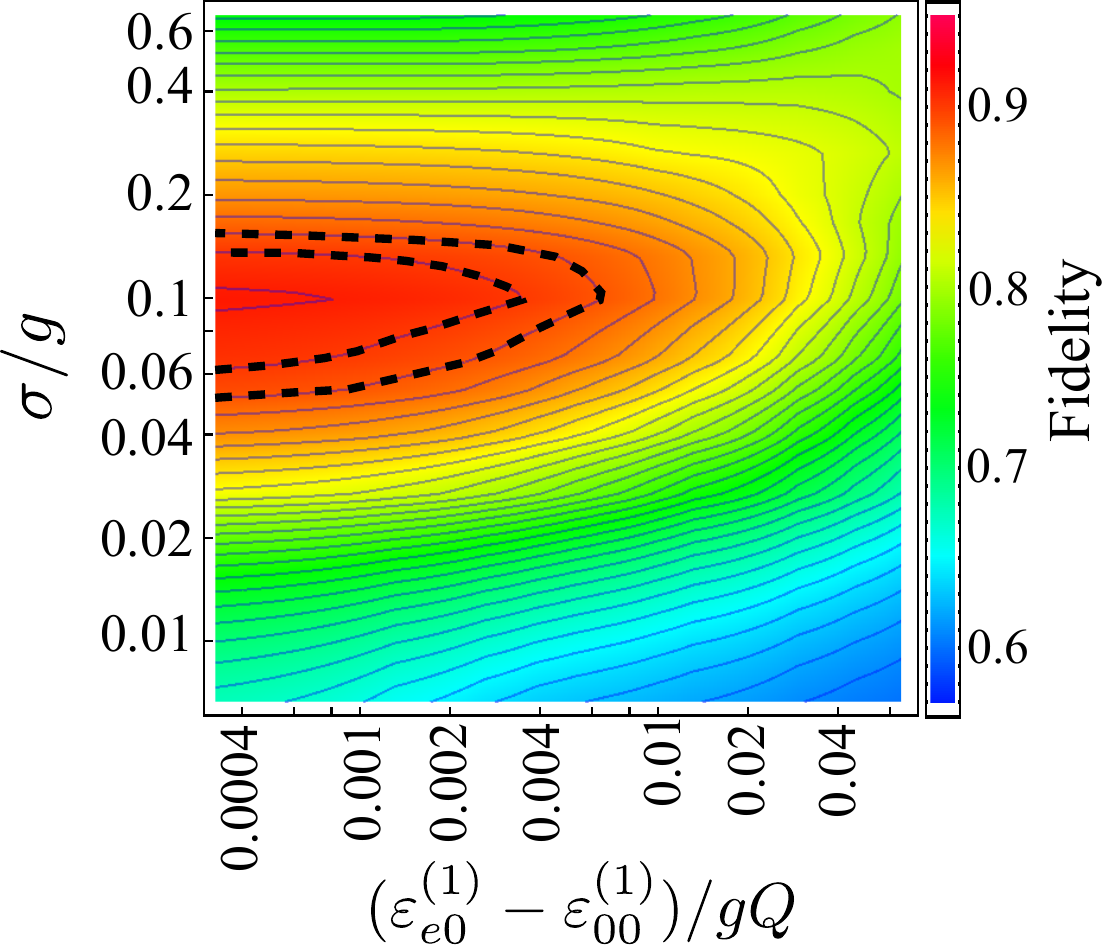}
\caption{\label{fig:NV:fidel-SQ}
Calculated fidelity of the two-qubit CZ gate performed in the system of two NV centers coupled via a microwave cavity mode.
Parameters are the same as in Fig.~\ref{fig:NV:N1}. The radiative and non-radiative decay rates are taken from experiment in Ref.~\onlinecite{Acosta}. Each point in the plot is optimized with respect to the position of $\delta$ (see the text and Ref. \onlinecite{solenov-NV} for details). The $90\%$ and $89\%$ fidelities are indicated by dashed curves (other contour curves are in steps of $1\%$).
}
\end{figure}

In order to estimate the fidelity of two-qubit operations in this system we compute the average fidelity (see Appendix~\ref{app:fidel}) numerically for the same CZ operation as earlier in Sec.~\ref{sec:QD}. In the case of the NV system, 
transitions $\LI{1}{B}=\LpI{1}{B}$ connect states $\v{00}$ and $\v{01}$ from the $\N=0$ subspace with states $\v{``e"0}$ and $\v{``e"1}$ from the $\N=1$ subspace, respectively. The energy curves corresponding to these states are highlighted in Fig.~\ref{fig:NV:N1}. The transition $\LII{2}{B}$ involves states $\v{``e"0}$ and $\v{``ee"}$ (the bottom-most energy curve in Fig.~\ref{fig:NV:N2}). With this choice of states the gate is expected to perform well for $\omega_C\gtrsim\varepsilon^{(1)}_{e0}$. As before, in the case of $\omega_C\lesssim\varepsilon^{(1)}_{e0}$ similar states to the left of the anti-crossings should be chosen to avoid populating the cavity mode.

In Fig. \ref{fig:NV:fidel-SQ} we show the averaged fidelity as a function of pulse bandwidth and cavity decay rate. As before, each point of the surface is optimized over the positioning of $\omega_C$. The fidelity shows the same dependence on parameters as for the InAs/GaAs quantum dot system, and it peaks at some value of pulse bandwidth $\sigma$ (same for all three pulses). At smaller values of $\sigma/g$ the pulses become slow and decoherence processes dominate reducing fidelity. At values of $\sigma/g$ approaching $1$ the pulses become insensitive to the energy shifts $\sim\omega_{\N=2}$ and later $\sim g$. As a result, accumulation of the phase at the chosen entangled state is hindered by involvement of other exited states reducing the fidelity.

\section{Superconducting transmon qubits}
\label{sec:SC}

\begin{figure}
\includegraphics[width=0.7\columnwidth]{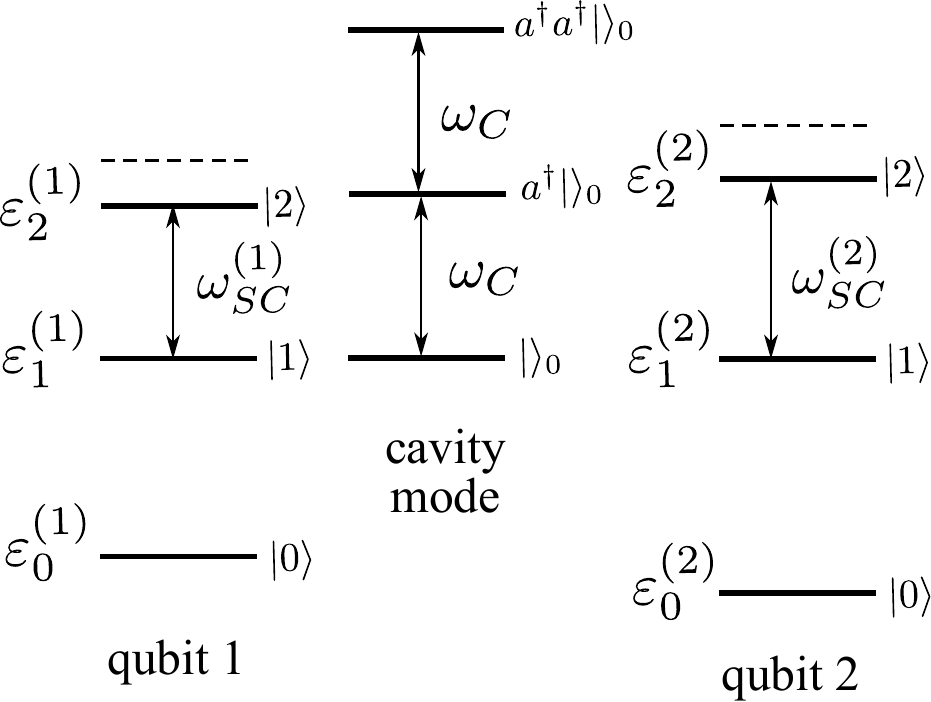}
\caption{\label{fig:SC:levels}
Two transmon qubits coupled to a microwave resonator. The resonator mode frequency $\omega_C$ is set near the $\omega^{(n)}_{SC}$ transition frequencies, with detuning $\delta=\omega_C-\omega^{(1)}_{SC}$. The first two states $\v{0}$ and $\v{1}$ of each transmon represent qubit states $\v{1}$ and $\v{0}$, respectively. The state $\v{2}$ is used as the auxiliary state $\v{e}$.   
}
\end{figure}

In this section we calculate the fidelity of the two-qubit CZ gate for a system of two superconducting transmon qubits\cite{SC-review,transmon} interacting with a microwave cavity mode.\cite{Lucero,Paik} The transmon qubit is a variation of a superconducting Cooper pair box system.\cite{SC-review} It includes two superconducting islands connected via one or two Josephson junctions. The system is assumed to be deep in the superconducting regime. In this case the low energy physics can be described by the Hamiltonian\cite{SC-review,Nigg}
\begin{eqnarray}\label{eq:SC:H}
H = 4E_C(\hat{{\frak n}}-{\frak n}_C)^2 - E_J \cos\hat{\phi},
\end{eqnarray}
where the first term is due to the capacitance, $C_{\rm eff}$, between the two superconducting islands, and the second term is due to the Josephson junction. The operators $\hat {\frak n}$ and $\hat\phi$ are particle number and phase operators, respectively;  $E_C = e^2/2C_{eff}$, and $E_J = \hbar I_c/2e$, where $I_c$ is the critical current through the junction. Transmons are typically tuned into a regime in which the first term is suppressed\cite{transmon,Nigg,SC-review} to reduce charge fluctuation noise ($E_C \ll E_J$). In this case the system can be well approximated by a harmonic oscillator with small anharmonicity,\cite{Nigg,SC-review}
\begin{eqnarray}\label{eq:SC:spectrum}
\varepsilon^{(n)}_N \approx (\omega^{(n)}_{SC}+3\eta^{(n)}) N - \eta^{(n)} N^2
\quad\quad
N = 0,1,2,...,
\end{eqnarray}
in which the qubit is encoded by the first two states $\v{0}$ and $\v{1}$ (see Fig.~\ref{fig:SC:levels}).
Both $\omega^{(n)}_{SC}$ and $\eta^{(n)}$ can be dynamically tuned.\cite{SC-review} In a system of two Josephson junctions per transmon,\cite{transmon} this is typically done by changing the magnetic flux $\Phi$ threading the loop between the junctions which alters the overall Josephson amplitude as $E_J = E^0_J\cos\Phi$, where $E^0_J$ is Josephson energy for each individual junction (if similar). Here we take 
\begin{eqnarray}\label{eq:SC:two-transmons}
\frac{\omega^{(2)}_{SC}}{\omega^{(1)}_{SC}}
=\frac{\eta^{(2)}}{\eta^{(1)}} = \chi,
\end{eqnarray}
and define $\Delta=\omega^{(2)}_{SC}-\omega^{(1)}_{SC} = \omega^{(1)}_{SC}(1-\chi)$ as before. Single qubit rotations in transmon systems are typically performed\cite{Lucero} by microwave pulses applied directly to the qubit states $\v{0}$ and $\v{1}$ (see Fig.~\ref{fig:SC:levels}). When the system is sufficiently anharmonic compared to the bandwidth of the pulses other transitions remain off resonance.

Two-qubit gates are typically performed\cite{Lucero} by coupling the transmon systems to a micro-strip or 3D microwave cavity mode. Because $\omega_{SC}$ can be dynamically tuned, both qubit transitions $(\varepsilon^{(1)}_1-\varepsilon^{(1)}_0)$ and $(\varepsilon^{(2)}_1-\varepsilon^{(2)}_0)$ can be simultaneously tuned in and out of resonance with the cavity mode, thus performing the entangling gate operations. Higher energy states either remain empty (slow passage) or have to be eliminated from the resulting evolution operator.

\begin{figure}
\includegraphics[width=0.99\columnwidth]{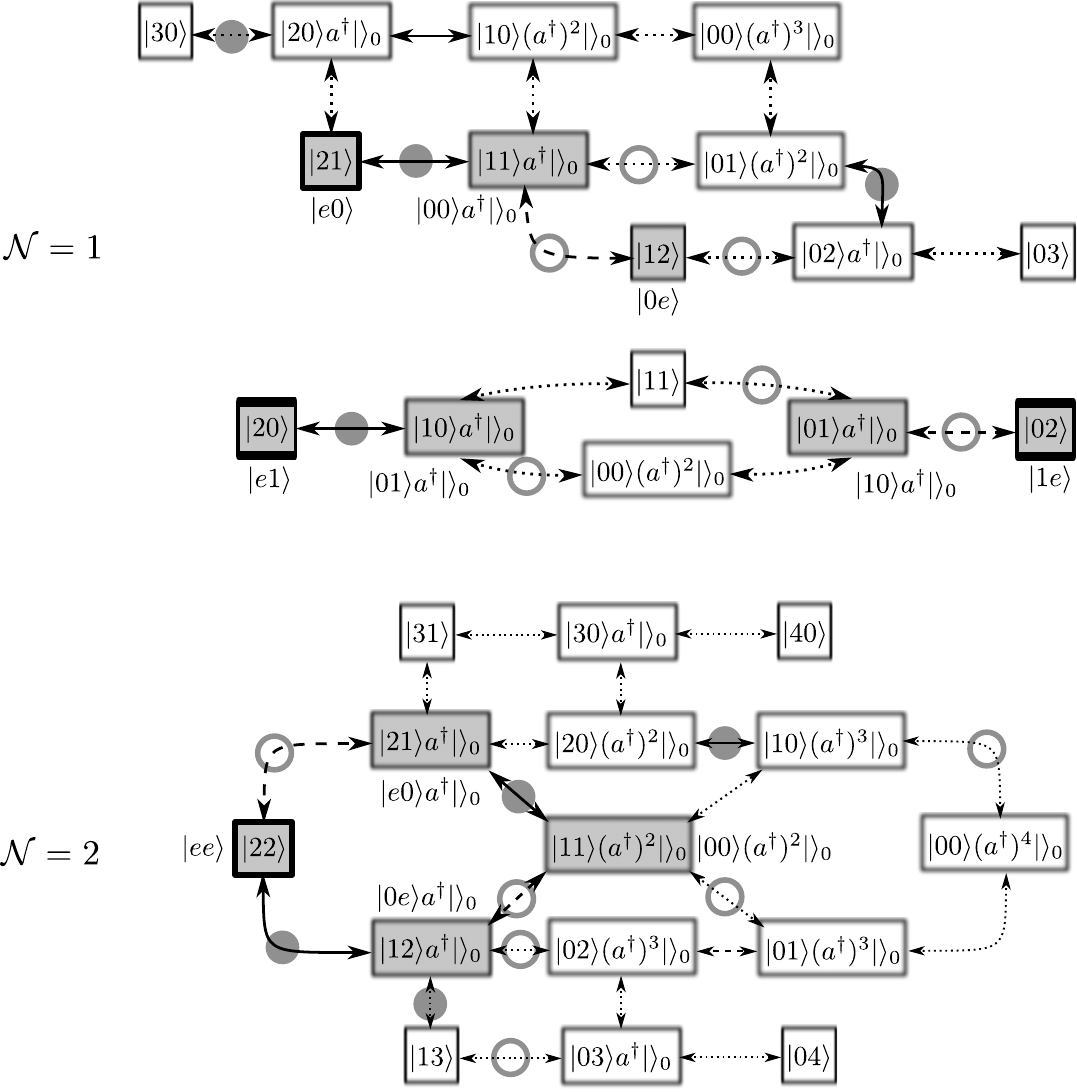}
\caption{\label{fig:SC:diagram}
The structure of the Hamiltonian for two transmon qubit systems coupled to a cavity mode. Subspaces involving states $\v{e0}$, $\v{e1}$, and $\v{ee}$ are shown. A state indicated in each box refers to transmon notation. The state labels outside of the boxes refer to standard notation introduced in Sec.~\ref{sec:2QC}. Solid and dashed connectors refer to cavity-induced coupling involving transition $\v{0}\leftrightarrow\v{e}$ in qubit-1 and qubit-2 systems, respectively. The cavity mode frequency $\omega_C$ is tuned close to these transitions. Cavity coupling to other transitions is shown by the dotted connectors. When $\eta\gg\Delta$, the latter coupling is small due to substantial detuning of $\omega_C$ from the corresponding energies. In this case, subspaces reduce to those typical for a simpler optical system discussed in Secs.~\ref{sec:QD} and \ref{sec:NV} (gray shade). Open and closed circles indicate comparable strengths of interactions for $2\eta\sim\Delta$ and $\delta\sim\Delta$: strong, $\sim g$, and weaker, $\sim g^2/\delta$, couplings, respectively.
}
\end{figure}

Dynamic tunability, however, often introduces an additional source of noise, thus reducing the coherence time of each qubit. In this case the pulse-controlled two-qubit gate operations introduced in Sec.~\ref{sec:2Q:gates} can become advantageous.
For better comparison with the earlier examples we will use the previously introduced $\v{1}$, $\v{0}$, and $\v{e}$ notation for states $\v{0}$, $\v{1}$, and $\v{2}$, respectively. The cavity is tuned to couple to the transition $\v{0}\leftrightarrow\v{e}$ as before ($\v{1}\leftrightarrow\v{2}$ in transmon notation). Unlike qubit systems in Secs.~\ref{sec:QD} and \ref{sec:NV}, transmons are nearly harmonic and higher and lower states can also participate. As a result, states $\v{e0}$, $\v{e1}$, and $\v{ee}$ couple to a large number of states when $g\neq 0$, as indicated in Fig.~\ref{fig:SC:diagram}. 

When $\eta\gg\Delta$, all transitions that are not $\v{0}\leftrightarrow\v{e}$ (dotted connectors in Fig.~\ref{fig:SC:diagram}) are suppressed. The coupling induced by the cavity in each $\N$ sector reduces to the one in the cavity-defect system described in Secs.~\ref{sec:QD} and \ref{sec:NV} (see also Appendix~\ref{app:interaction} and Fig.~\ref{fig:app:A:diagram}). The important difference, however, is that the systems described in Secs. \ref{sec:QD} and \ref{sec:NV} had two auxiliary states $\v{e_0}$ and $\v{e_1}$ coupled to one another indirectly, whereas in the superconducting qubit system only one $\v{e}$ is available in this limit [see Figs.~\ref{fig:2QC-cartoon}(b) and \ref{fig:SC:levels}]. This affects the values of $\Delta\omega_{\N=1}$ and $\Delta\omega_{\N=2}$. We can use the derivation for the system with two auxiliary states given in Appendix~\ref{app:interaction} and set $\delta\omega \sim \varepsilon_{e_1}-\varepsilon_{e_0} \to \infty$ thereby removing state $\v{e_1}$ (cf. Fig.~\ref{fig:app:A:diagram} and the highlighted states in Fig.~\ref{fig:SC:diagram}). In this case we obtain $\Delta\omega_{\N=1}\sim \Delta\omega_{\N=2}\sim g^4/\Delta^3$ and the intermediate regime of resonance defined in Sec.~\ref{sec:2Q:inter} is not accessible.

When $\eta\sim\Delta$, states $\v{ee}$, $\v{e0}$, and $\v{0e}$ are coupled to a larger number of states. Specifically, the interactions represented by dotted connectors (Fig.~\ref{fig:SC:diagram}) and marked with closed or open circles become comparable with the interactions between the shaded states (solid or dashed connectors). In this case the intermediate regime of resonance $\Delta\omega_{\N=1}/\Delta\omega_{\N=2}\ll 1$ can be achieved. To demonstrate this, we analyze the case $2\eta\approx\Delta$ (i.e., $2\eta-\Delta\sim g$) and estimate 
$\Delta\omega_{\N=1}$ and $\Delta\omega_{\N=2}$.

The energy difference $\Delta\omega_{\N=1}$ is based on energies $E_{``e"0}$ and $E_{``e"1}$ ($E_{``2"1}$ and $E_{``2"0}$ in transmon notations). State $\v{21}$ interacts weakly ($\sim g^2/\Delta$; closed circles in Fig.~\ref{fig:SC:diagram}) with state $\v{11}a^\dag\v{}_0$. The latter, in turn, interacts strongly ($\sim g$; open circles in Fig.~\ref{fig:SC:diagram}) with three other states: $\v{12}$, $\v{01}(a^\dag)^2\v{}_0$, and (indirectly) $\v{02}a^\dag\v{}_0$. Therefore, for $\delta\to\Delta$, the energy $E_{``2"1}$ is shifted by $\approx -(1/4)\sum_i g^2/(\Delta+\epsilon_i)$, where $\epsilon_i$ are energies of the four states mentioned above relative to $E_{21}+\Delta$. The energies $\epsilon_i$ are distributed symmetrically with respect to their average $\bar\epsilon = \sum_i\epsilon_i/4$. We obtain
\begin{eqnarray}\label{eq:SC:E21}
E_{``2"1} = E_{21} - \frac{g^2}{\Delta} + \frac{g^2 \bar\epsilon}{\Delta^2} + \O(g^4/\Delta^3).
\end{eqnarray}
The state $\v{20}$ is weakly coupled to state $\v{10}a^\dag\v{}_0$ that is strongly coupled to $\v{00}(a^\dag)^2$. Therefore the energy $E_{``2"0}$ can be estimated similarly, except that only two energies enter the sum as $\epsilon'_i$. The average $\bar\epsilon = \sum_i\epsilon'_i/2$ is the same and we obtain
\begin{eqnarray}\label{eq:SC:E20}
E_{``2"0} = E_{20} - \frac{g^2}{\Delta} + \frac{g^2 \bar\epsilon}{\Delta^2} + \O(g^4/\Delta^3).
\end{eqnarray}
Note, however, that the terms $\sim 1/\Delta^3$ in Eqs.~(\ref{eq:SC:E21}) and (\ref{eq:SC:E20}) are different, since $\sum_i\epsilon_i^{\prime 2}/2\neq \sum_i\epsilon_i^2/4$, but both $\sim g^4$. When $\delta > \Delta$, the first- and the second-order terms ($\sim 1/\delta$ and $\sim 1/\delta^2$ in this case) are still the same for both $E_{``2"0}$ and $E_{``2"1}$, and we obtain 
\begin{eqnarray}\label{eq:SC:N1}
\Delta\omega_{\N=1} \lesssim g^4/\Delta^3.
\end{eqnarray}
Note that the region $\delta-\Delta\sim g$ has to be excluded to avoid strong interaction between states $\v{02}$ and $\v{11}$. This interference drives the system away from the intermediate resonance regime. The result (\ref{eq:SC:N1}) is similar to the one obtained for the defect-cavity systems in Appendix~\ref{app:interaction}. In the latter case, however, the terms $\sim\Delta^3$ are also canceled due to small $\delta\omega$ and $\Delta\omega_{\N=1}$ is reduced by $\delta\omega/\Delta$.

The energy difference $\Delta\omega_{\N=2}$ can be estimated based on the energies of states $\v{ee}$, $\v{e0}$, and $\v{1e}$ (states $\v{22}$, $\v{21}$, and $\v{02}$ in transmon notations respectively). When $\delta>\Delta$ and we stay away from the region $\delta-\Delta\sim g$, the energy corresponding to state $\v{02}$ is
\begin{eqnarray}\label{eq:SC:E02}
E_{0``2"} = E_{02} - 2g^2/\delta + \O(1/\delta^3).
\end{eqnarray}
The state $\v{22}$ interacts strongly with state $\v{21}a^\dag\v{}_0$ and weakly with state $\v{12}a^\dag\v{}_0$. Both links, however, are suppressed due to non-zero detuning $\delta-\Delta$. The energy corresponding to the state $\v{22}$ can be found as
\begin{eqnarray}\label{eq:SC:E22}
E_{``ee"}-E_{ee} \sim - g/\delta.
\end{eqnarray}
Note that the proportionality coefficient in this case is not 2 as it is in Eq.~(\ref{eq:SC:E02}). Therefore we obtain
\begin{eqnarray}\label{eq:SC:E22}
\Delta\omega_{\N=2} \sim g/\delta,
\end{eqnarray}
and
\begin{eqnarray}\label{eq:SC:N1N2}
\Delta\omega_{\N=1}/\Delta\omega_{\N=2} \sim g^2/\Delta^2,
\end{eqnarray}
for $\delta\gtrsim\Delta$. This is larger by a factor of $\Delta/\delta\omega$ than $\Delta\omega_{\N=1}/\Delta\omega_{\N=2}$ achievable in the defect-cavity systems discussed in Secs.~\ref{sec:QD} and \ref{sec:NV}. In the case of a superconducting qubit-cavity system the qubit-subspace states play the role of the extra auxiliary state $\v{e_1}$ and $\delta\omega$ remains effectively $\sim\Delta$. The result (\ref{eq:SC:N1N2}) applies to $|2\eta-\Delta|$ up to $\sim\Delta$ (and, possibly, even further in some cases) as can be verified by numerical modeling of the entangling gate operations in this system.

\begin{figure}
\includegraphics[width=0.8\columnwidth]{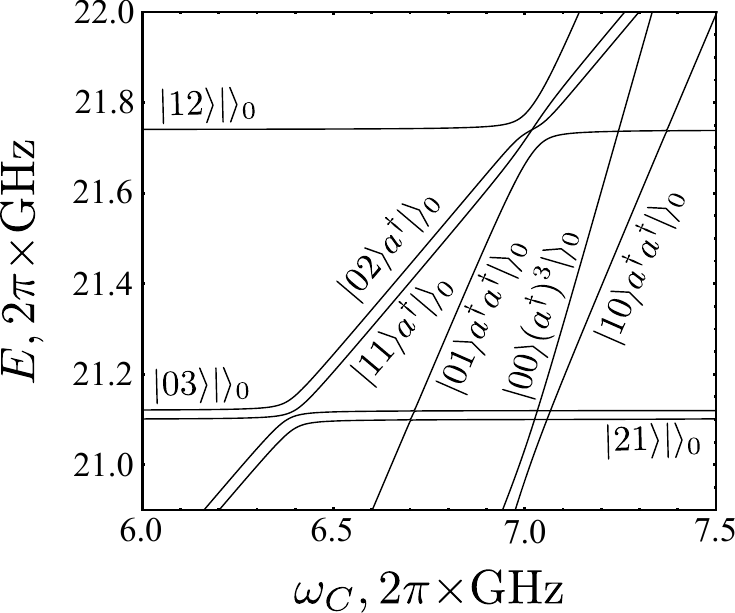}
\caption{\label{fig:SC:N1}
Energies of the $\N=1$ sector states as a function of the cavity frequency for the system of two transmons coupled to a microwave cavity. The parameters are
$\omega^{(1)}_{SC} = 6.4\times 2\pi$ GHz, $\eta^{(1)} = 0.3\times 2\pi$ GHz,  $\chi = 1.1$ ($\Delta=0.64\times 2\pi$ GHz), and $g = 20\times 2\pi$ MHz. Note that $g\ll\Delta\sim\eta$.
}
\end{figure}

As before, we focus on the $CZ={\rm diag}\{1,-1,1,1\}$ gate outlined in Eq.~(\ref{eq:2Q:CU-pulses-eg3}). Transitions $\LI{1}{B}=\LpI{1}{B}$ connect states $\v{11}$ and $\v{10}$ (in transmon notations) from the $\N=0$ subspace with states $\v{``2"1}$ and $\v{``2"0}$ from the $\N=1$ subspace, respectively (see Fig.~\ref{fig:SC:N1}). The transition $\LII{2}{B}$ connects states $\v{``2"1}$ and $\v{``22"}$ (see Fig.~\ref{fig:SC:N2}).

In Fig.~\ref{fig:SC:fidel} we plot the average fidelity (see Appendix~\ref{app:fidel}) of the CZ gate calculated numerically with the cavity mode decoherence rate $\Gamma_C = 7.6\times 10^{-5}\times 2\pi$ GHz ($Q\sim 10^5$) and the qubit decoherence rate $\Gamma = 4.2\times 10^{-6}\times 2\pi$ GHz (decoherence time is $\sim 40\mu$s), as reported in Ref.~\onlinecite{Paik} for a transmon in a 3D cavity. The fidelity has a wide plateau of high values ($\gtrsim 98\%$) for $\delta=\omega_C-\omega^{(1)}_{SC}$ between $\Delta$ and $2\Delta$. It is suppressed at $\delta-\Delta\sim g$ due to interference with transition $\v{0}\leftrightarrow\v{1}$ in the first qubit, which results in a large value of $\Delta\omega_{\N=1}$ as explained before. At $\delta-2\Delta\sim g$ the fidelity is reduced due to interference with transition $\v{0}\leftrightarrow\v{1}$ in the second qubit. At small values of $\sigma$, the duration of the pulses become comparable with the decoherence times and the fidelity is suppressed as well. At values of $\sigma\gtrsim g$ the pulses become too broad to resolve the interaction. At larger values of $\delta$ the magnitude of $\Delta\omega_{\N=2}$ becomes comparable with the decoherence rates and entangling gates are no longer possible.

\begin{figure}
\includegraphics[width=0.99\columnwidth]{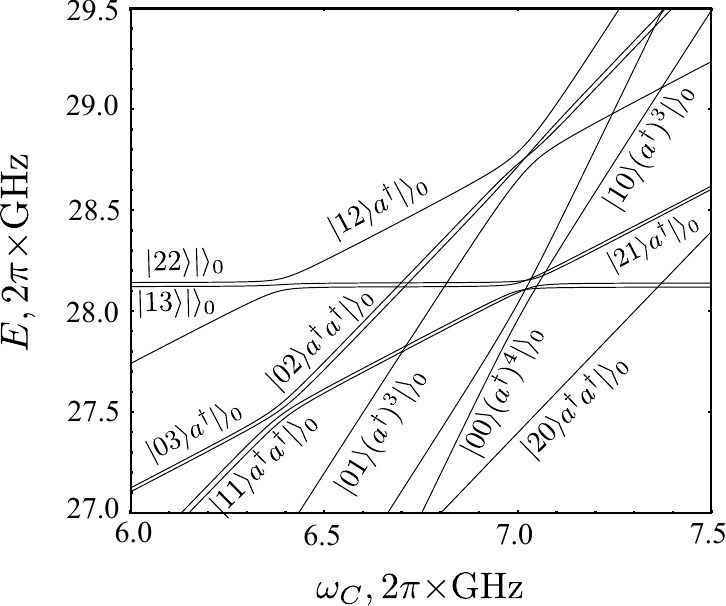}
\caption{\label{fig:SC:N2}
Energies of the $\N=2$ sector states as a function of the cavity frequency for a system of two transmons coupled to a microwave cavity. The parameters are the same as in Fig.~\ref{fig:SC:N1}.
}
\end{figure}
\begin{figure}
\includegraphics[width=0.99\columnwidth]{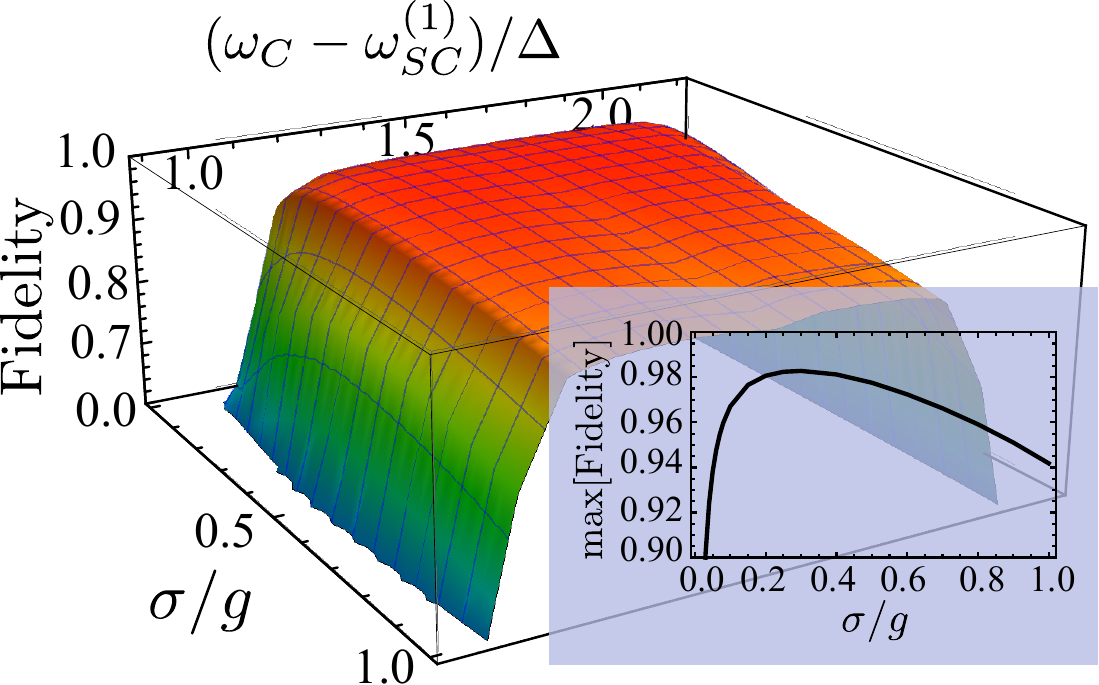}
\caption{\label{fig:SC:fidel}
Calculated average fidelity of the CZ gate in the system of two transmons coupled to a microwave cavity. The fidelity is shown as a function of the cavity frequency, $\omega_C$, and the bandwidth of the pulses $\sigma$. Inset shows average fidelity optimized over $\omega_C$. The parameters are the same as in Fig.~\ref{fig:SC:N1}.
}
\end{figure}

\section{Conclusions}

We investigated a two-qubit system interacting via a bosonic cavity mode. We demonstrated that auxiliary excited states in each qubit system can be used to perform two-qubit entangling operations efficiently. A generalized approach that encompasses a number of physical systems was developed. We showed that two multi-state quantum systems interacting via a cavity mode can develop an intermediate regime of resonance in which some types of auxiliary states remain local to each qubit system while others become non-local. In this regime single-qubit pulse controls that involve auxiliary states do not generally commute. This effect is the basis for pulse-controlled entangling operations that can coexist with similarly performed fast single-qubit operations, and do not require dynamical tunability of states. Examples of three physical systems were given: self-assembled quantum dot qubits, NV centers in diamond, and superconducting transmon qubits. 

In all systems discussed an intermediate resonance regime can be achieved when transitions to excited states are sufficiently different compared to qubit-cavity coupling. In a fully coherent system the gates based on the intermediate resonance regime can be performed with arbitrarily high precision, with fidelities reaching 100\%. In an actual physical system fidelity is always limited by decoherence. In the examples discussed in Secs.~\ref{sec:QD} (quantum dots) and~\ref{sec:NV} (NV-centers) the fidelity was computed for the realistic values of decoherence dominated by that of the cavity mode. In these systems higher cavity mode quality factors are necessary to reach fidelities needed for sustainable quantum computation (over 99\%). While for the superconducting transmon systems (Sec.~\ref{sec:SC}) the calculated fidelities (of up to $\approx 98\%$) were also limited primarily by the cavity mode quality factor ($Q\sim 10^5$), it should be noted that the calculations were performed for decoherence values reported for a transmon in a 3D cavity\cite{Paik}. For currently available transmission line cavities substantially higher quality factors and larger values for qubit-cavity coupling, $g$, can be reached. Since the time of a two-qubit gate operation scales as $1/g$ and qubit decoherence rates are about one order of magnitude smaller that that of a 3D cavity mode used in this work, fidelity rates in excess of 99\% are expected to be achievable. This makes superconducting transmon-based systems currently the most promising candidates to implement and benefit from the approach developed in this paper. At the same time, unlike in quantum dots or NV-center systems, in superconducting transmon systems the excitation ``bands" that define the gate operations (sketched in Fig.~\ref{fig:2QC:spec}) are not separated, but, in fact, overlap in energy. This makes the visualization of the developed approach more difficult. In addition, the existence of the intermediate regime of resonance becomes dependent on the value of anharmonicity in transmon systems.

This work was supported in part by the ONR, NRC/NRL, and LPS/NSA. Computer resources were provided by the DOD HPCMP.

\appendix

\section{Degree of interaction in one- and two-excitations sector}\label{app:interaction}

\begin{figure}
\includegraphics[width=0.99\columnwidth]{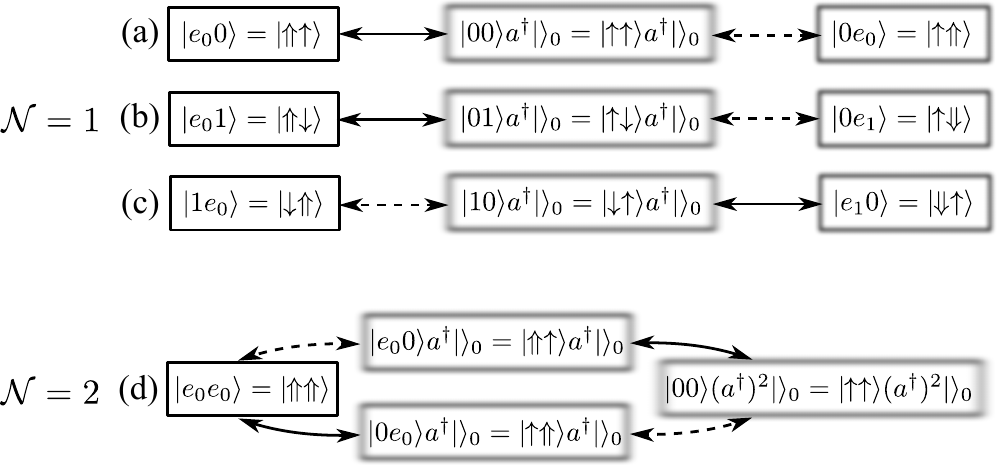}
\caption{\label{fig:app:A:diagram}
The structure of the Hamiltonian $H_0$ for subspaces involving states $\v{e_00}$, $\v{e_01}$, $\v{1e_0}$, and $\v{e_0e_0}$. Solid connectors indicate cavity coupling to the first qubit system and dashed connectors represent cavity coupling to the second qubit system.
}
\end{figure}

We consider the first two excited states for each qubit system and use spin notation, i.e., $\v{\U}\equiv\v{e_1}$, $\v{\D}\equiv\v{e_2}$, $\v{\u}\equiv\v{0}$, and $\v{\d}\equiv\v{1}$. Since the number of excitations $\N$ is conserved, the states in the Hamiltonian $H_0$ in Eq.~\ref{eq:2QC:H0} can be grouped into closed subsets. Several such subsets corresponding to $\N=1$ and $\N=2$ are shown in Fig.~\ref{fig:app:A:diagram}. Coupling to the cavity changes the energies and mixes these states. Changes in $\N=1$ states are typically substantially different from those in $\N=2$ states. This stems from the different structure of interaction (connectivity) in $\N=1$ and $\N=2$ subsets (see Fig.~\ref{fig:app:A:diagram}). Here we analyze the states that are important for two-qubit gate operations. We first look at $\Delta\omega_{\N=1} = \omega_{\u\d\leftrightarrow\U\d}-\omega_{\u\u\leftrightarrow\U\u}$, where $\omega_{i\leftrightarrow j} = E_j-E_j$, see Eqs.~(\ref{eq:2QC:omega-N1}) and (\ref{eq:2QC:H0-spectrum}), and the states are labeled by the dominant spin configuration for clarity. This transition energy difference describes the effect of the state of the second qubit on transitions involving the first qubit. When $\Delta\omega_{\N=1}=0$, transitions between $\N=0$ and $\N=1$ states are completely local to each qubit system. We also look at 
$\Delta\omega_{\N=2} = \omega_{\d\u\leftrightarrow\d\U}-\omega_{\U\u\leftrightarrow\U\U}$. This difference is similar to $\Delta\omega_{\N=1}$, but involves one $\N=2$ state. When $\Delta\omega_{\N=2}=0$, operations involving $\N=2$ states are also local and do not depend on the state of the other qubit system. In what follows we demonstrate that the dependence of $\Delta\omega_{\N=1}$ on $\Delta$ can be substantially different from that of $\Delta\omega_{\N=2}$ in the limit of large $\Delta$.

\subsection{$\N=1$ sector}
The difference $\Delta\omega_{\N=1}$ depends on the energies $E_{\u\u}$, $E_{\u\d}$, $E_{\U\u}$, and $E_{\U\d}$. The states $\v{\U\u}$ and $\v{\U\d}$ are parts of two closed subsets of states, Fig.~\ref{fig:app:A:diagram}(a) and Fig.~\ref{fig:app:A:diagram}(b), respectively. The energy $E_{\U\u}$ is the bottom branch in the spectrum of the Hamiltonian projected onto states $\v{\U,\u}$, $\v{\u,\U}$, and $\v{\u\u}a^\dag\v{}_0$ [see Fig.~\ref{fig:app:A:diagram}(a)], 
\begin{eqnarray}\label{eq:app:int:N1:Huu}
H_{\U\u} = 
\left(
\begin{array}{cccc}
0 & 0 & g\\
0 & \Delta & g\\
g & g & \delta \\
\end{array}
\right).
\end{eqnarray}
Here the energy was shifted by $\varepsilon^{(1)}_{\u\u}$ and we recall that $\delta = \omega_C-(\varepsilon^{(1)}_\U-\varepsilon^{(1)}_\u)$ and $\Delta = (\varepsilon^{(2)}_\U-\varepsilon^{(1)}_\U)-(\varepsilon^{(2)}_\u-\varepsilon^{(1)}_\u)$. Similarly, the energy $E_{\U\d}$ is the bottom branch in the spectrum of the Hamiltonian projected onto states $\v{\U,\d}\v{}_0$, $\v{\u,\D}\v{}_0$, and $\v{\u\d}a^\dag\v{}_0$ [see Fig.~\ref{fig:app:A:diagram}(b)],
\begin{eqnarray}\label{eq:app:int:N1:Hud}
H_{\U\d} = 
\left(
\begin{array}{cccc}
0 & 0 & g\\
0 & \Delta + \delta\omega & g\\
g & g & \delta \\
\end{array}
\right),
\end{eqnarray}
where the energy was shifted by $\varepsilon^{(1)}_{\u\d}$, and $\delta\omega = (\varepsilon^{(n)}_\U-\varepsilon^{(n)}_\D) - (\varepsilon^{(n)}_\u-\varepsilon^{(n)}_\d)$ is assumed to be the same for both qubit systems for clarity. Note that $H_{\U\d}(\Delta,\delta\omega) = H_{\U\u}(\Delta+\delta\omega)$, $E_{\u\u}=\varepsilon_{\u\u}$, and $E_{\u\d}=\varepsilon_{\u\d}$. Therefore we obtain
\begin{eqnarray}\label{eq:app:int:N1:omega1-def}
\Delta\omega_{\N=1} &=&
[E_{\U\d}-E_{\u\d}]-[E_{\U\u}-E_{\u\u}]
\\\label{eq:app:int:N1:omega1-def-small-dw}
&\xrightarrow[]{\delta\omega\ll\Delta}&
\delta\omega \times \partial_{\Delta}E_{\U\u}(\Delta)
\\\label{eq:app:int:N1:omega1-def-large-dw}
&\xrightarrow[]{\delta\omega\to\infty}&
E_{\U\u}(\infty)-E_{\U\u}(\Delta)
\end{eqnarray}
Note that the sign of $\delta\omega$ is not important for the derivations, and we set $\delta\omega>0$.
The spectrum of (\ref{eq:app:int:N1:Huu}) is given by
\begin{eqnarray}\label{eq:app:int:N1:det}
E(E-\delta)(E-\Delta) = g^2(2E-\Delta),
\end{eqnarray}
which can be solved exactly. It is instructive, however, to obtain the asymptotic behavior for $E_{\U\u}(\Delta)$. When $\Delta\to\infty$ and the cavity is in resonance with the $\v{\u\u}\leftrightarrow\v{\U\u}$ transition, i.e., $\delta=0$, the interaction with the cavity splits the degeneracy between states $\v{\U\u}\v{}_0$ and $\v{\u\u}a^\dag\v{}_0$, and the lowest two energies of the spectrum are $\pm g$. When $\Delta$ is finite (non-zero) and positive, the interaction with the state $\v{\u\U}\v{}_0$ shifts these energies down.

\subsubsection{the limit of $\delta\to 0$}
When $\delta=0$ the energy $E_{\U\u}(\Delta)$ is defined by
\begin{eqnarray}\label{eq:app:int:N1:E-0}
E^2 = g^2 - g^2 \frac{E}{\Delta - E},
\end{eqnarray}
and
\begin{eqnarray}\label{eq:app:int:N1:dE-0}
\partial_\Delta E = \frac{1}{2E}\partial_\Delta E^2 
= \frac{1}{2}\frac{g^2}{(\Delta - E)^2} 
\xrightarrow[]{g\ll\Delta}
\frac{1}{2}\frac{g^2}{\Delta^2}.
\end{eqnarray}
As a result, we obtain 
\begin{eqnarray}\label{eq:app:int:N1:omega1-0}
\Delta\omega_{\N=1} \xrightarrow[\delta\omega\ll\Delta]{\delta=0,\,\,g\ll\Delta} 
\frac{1}{2}\frac{g^2\delta\omega}{\Delta^2},
\end{eqnarray}
and
\begin{eqnarray}\label{eq:app:int:N1:omega1-0-infDW}
\Delta\omega_{\N=1} \xrightarrow[\delta\omega\to\infty]{\delta=0,\,\,g\ll\Delta} 
\frac{1}{2}\frac{g^2}{\Delta}.
\end{eqnarray}

\subsubsection{the case of $\delta\gtrsim\Delta$}
In this case, the bottom branch of the spectrum is given by 
\begin{eqnarray}\label{eq:app:int:N1:E-inf}
E = g^2\frac{2E-\Delta}{(E-\delta)(E-\Delta)},
\end{eqnarray}
where the right-hand side should be computed iteratively. The zeroth order iteration ($E\to 0$) gives $-g^2/\delta$ which does not depend on $\Delta$. Expanding the denominator in the powers of $E$ and collecting the orders of $g^2$ after two iterations we obtain 
\begin{eqnarray}\label{eq:app:int:N1:dE-inf}
E = - \frac{g^2}{\delta} - \frac{g^4}{\delta^2}
\left( \frac{1}{\Delta} - \frac{1}{\delta} \right)
+\O(g^6/\delta^3\Delta^2).
\end{eqnarray}
After a straightforward differentiation we obtain
\begin{eqnarray}\label{eq:app:int:N1:omega1-inf}
\Delta\omega_{\N=1} 
\xrightarrow[\delta\omega\ll\Delta]{\delta\gtrsim\Delta,\,\,g\ll\Delta} 
\frac{g^4\delta\omega}{\Delta^2\delta^2},
\end{eqnarray}
and
\begin{eqnarray}\label{eq:app:int:N1:omega1-inf-infDW}
\Delta\omega_{\N=1} 
\xrightarrow[\delta\omega\to\infty]{\delta\gtrsim\Delta,\,\,g\ll\Delta} 
\frac{g^4}{\Delta\delta^2}.
\end{eqnarray}
The limits (\ref{eq:app:int:N1:omega1-0}) and (\ref{eq:app:int:N1:omega1-inf}) together with the full numerical solution are shown in Fig.~\ref{fig:2QC:int}.

\subsection{$\N=2$ sector}

The difference $\Delta\omega_{\N=2}$ depends on the energies $E_{\d\u}$, $E_{\d\U}$, $E_{\U\u}$, and $E_{\U\U}$. The energies $E_{\d\U}$ and $E_{\U\u}$ are found by analyzing the Fig.~\ref{fig:app:A:diagram}(a) and Fig.~\ref{fig:app:A:diagram}(c) subsets of states as in the previous subsection. The energy $E_{\U\U}$ is found by diagonalizing subset (d) in Fig.~\ref{fig:app:A:diagram}. The energy $E_{\U\U}$ is the bottom branch of the spectrum of the Hamiltonian projected onto states $\v{\U\u}a^\dag\v{}_0$, $\v{\u\U}a^\dag\v{}_0$, $\v{\u\u}a^\dag a^\dag\v{}_0$, and $\v{\U\U}$, 
\begin{eqnarray}\label{eq:app:int:N2:Huu}
H_{\U\U} = 
\left(
\begin{array}{cccc}
\delta-\Delta & 0 & g & g\\
0 & \delta & g & g\\
g & g & 2\delta-\Delta & 0 \\
g & g & 0 & 0\\
\end{array}
\right),
\end{eqnarray}
where we have shifted the energy by $\varepsilon^{ee}_{\u\u}$.
The spectrum of (\ref{eq:app:int:N2:Huu}) is given by 
\begin{eqnarray}\label{eq:app:int:N2:det}
(\!E\!-\!\delta\!+\!\Delta\!)(\!E\!-\!\delta)(\!E\!-\!\!2\delta\!+\!\Delta\!)E \!=\! g^2(2\delta\!-\!\Delta\!-\!\!2E)^2,
\end{eqnarray}
and can be found exactly. As before, it is instructive to obtain the limit $\delta \to \Delta$, with $g\ll\Delta$.

\subsubsection{The limit of $\delta\to\Delta$}
In this case, the lowest branch of the spectrum of Eq.~(\ref{eq:app:int:N2:Huu}) is given by the bottom energy branch, $E<0$, of
\begin{eqnarray}\label{eq:app:int:N2:det-low}
(E-\Delta)E = g(\Delta-2E),
\end{eqnarray}
and we find
\begin{eqnarray}\nonumber
E_{\U\U}-\varepsilon^{ee}_{\u\u} 
&=& 
\frac{\Delta}{2}-g-\sqrt{\frac{\Delta^2}{4}+g^2}
\\\label{eq:app:int:N2:EUU}
&=& 
-g -\frac{g^2}{\Delta} + \frac{g^4}{\Delta\!^3} + \O(g^6/\Delta^5).
\end{eqnarray}
The energy $E_{\U\u}$ is the bottom branch of the spectrum of Eq.~(\ref{eq:app:int:N1:det}) and can be found from 
\begin{eqnarray}\label{eq:app:int:N2:Eeq-Uu}
E = -\frac{g^2}{\Delta-E} +\frac{g^2E}{(\Delta-E)^2}
\end{eqnarray}
iteratively. After one iteration we obtain
\begin{eqnarray}\label{eq:app:int:N2:EUu}
E_{\U\u}-\varepsilon^{(1)}_{\u\u} = 
-\frac{g^2}{\Delta}+\frac{g^6}{\Delta^5}+\O(g^8/\Delta^7).
\end{eqnarray}
Therefore we have
\begin{eqnarray}\label{eq:app:int:N2:wUU}
\omega_{\U\u\leftrightarrow\U\U} - \varepsilon^{(2)}_{\u\u}
=
-g+\frac{g^4}{\Delta^3}+\O(g^6/\Delta^5).
\end{eqnarray}
Note that the $g^2/\Delta$ term cancels out.

The transition energy $\omega_{\d\u\leftrightarrow\d\U}$ can be found from the $E\to\Delta$ branch of the Hamiltonian (\ref{eq:app:int:N1:Hud}) with the substitution $\Delta\to\Delta-\delta\omega$, $\delta\to\delta-\delta\omega$, and the additional total energy shift of $\delta\omega$. When $\delta=\Delta$ it can be found iteratively from
\begin{eqnarray}\label{eq:app:int:N2:Eeq-du}
E-\Delta = -g \sqrt{1+\frac{E-\Delta}{\Delta+(E-\Delta)}}. 
\end{eqnarray}
After three iterations we find
\begin{eqnarray}\label{eq:app:int:N2:wdu}
\omega_{\d\u\leftrightarrow\d\U} = 
-g+\frac{1}{2}\frac{g^2}{\Delta}-\frac{7}{8}\frac{g^3}{\Delta^2} + \O(g^4/\Delta^3).
\end{eqnarray}

When $\delta\omega\sim g\ll\Delta$, the $\delta\omega$ shifts can be neglected. Subtracting Eq.~(\ref{eq:app:int:N2:wUU}) from Eq.~(\ref{eq:app:int:N2:wdu}) we obtain
\begin{eqnarray}\label{eq:app:int:N2:omega2-0}
\Delta\omega_{\N=2} 
\xrightarrow[]{\delta=\Delta,\,\,g\ll\Delta} 
\frac{1}{2}\frac{g^2}{\Delta}.
\end{eqnarray}
This limit together with the full numerical solution is shown in Fig.~\ref{fig:2QC:int}. When $\delta\omega\to\pm\infty$ we have to replace $\Delta\to\delta\omega$ in Eq.~(\ref{eq:app:int:N2:wdu}). In this case we obtain $\omega_{\d\u\leftrightarrow\d\U} = -g$ and  $\Delta\omega_{\N=2}\sim g^4/\Delta^3$.

\section{Phases due to multiple pulses}\label{app:phase-lock}

When a quantum gate operation is diagonal and involves a sequence of non-overlapping pulses that, by the end of the sequence, restores the system to the qubit subspace, phases associated with each pulse field do not enter the result. To demonstrate this, consider the case of three pulses, $\{\pi,U_0,\pi\}$, that involve excited states $\v{b^{(n)}_j}$ and qubit sub-space states $\v{a^{(n)}_j}$. Here the upper index refers to the qubit system, and the lower index refers to a specific state of that system. The first and the last pulses are identical $\pi$ (population inversion or swap) pulses. The middle pulse $U_0$ is different. The gate is performed in a rotating frame of reference, with the total evolution operator given by
\begin{eqnarray}\label{eq:app:pulses:U}
U_{tot} = e^{-iH_0(t_f-t_i)} U(t_f,t_i),
\end{eqnarray}
with
\begin{eqnarray}\label{eq:app:pulses:Ut}
U(t_f,t_i) = 
T\exp\left[
\int_{t_i}^{t_f}dt' e^{iH_0t'}V(t')e^{-iH_0t'}
\right].
\end{eqnarray}
We will assume that the pulses do not overlap in time significantly. In this case
\begin{eqnarray}\label{eq:app:pulses:U}
U(t_f,t_i) = U_3(t_f,t_i)U_2(t_f,t_i)U_1(t_f,t_i).
\end{eqnarray}
The $\pi$ pulses are resonant pulses. They swap the population between the states. Typically there are several transitions that can be affected by each $\pi$ pulse. For example, if the interaction through the cavity is ineffective (or absent) each transition out of the qubit subspace of a two-qubit system is two-fold degenerate: transitions such as $\v{00}\leftrightarrow\v{e0}$ and $\v{01}\leftrightarrow\v{e1}$ are identical. Therefore $U_1$ and $U_3$ are 
\begin{eqnarray}\label{eq:app:pulses:pi}
U_\pi = \prod_j
\left[
(-i)\left(
\v{a^n_j}\iv{b^n_j}e^{-i\phi^n_{a_jb_j}}
+h.c.
\right)
\right],
\end{eqnarray}
where $\phi^n_{a_jb_j}$ is the sum of the phase of the pulse field $\phi_p$ and the phase of the interaction matrix element [see Eq.~\ref{eq:2QC:V}]. These two pulses bring the population out of the qubit subspace and return it back. If $U_0$ keeps the population on $\v{b^n_j}$ states after the pulse and does not affect any of the $\v{a^n_j}$ states, it can be formulated as
\begin{eqnarray}\label{eq:app:pulses:U0}
U_0 = \sum_{ll'}B_{ll'}\v{b^n_l}\iv{b^n_{l'}},
\end{eqnarray}
where $B_{ll'}$ are some complex coefficients.
Different terms of Eq.~(\ref{eq:app:pulses:U0}) and Eq.~(\ref{eq:app:pulses:pi}) can be grouped into sequences such as
\begin{eqnarray}\label{eq:app:pulses:seq}
(-i)\v{a^n_l}\iv{b^n_l}e^{-i\phi^n_{a_lb_l}}
B_{ll'}\v{b^n_l}\iv{b^n_l}
(-i)\v{b^n_{l'}}\iv{a^n_{l'}}e^{i\phi^n_{a_{l'}b_{l'}}}\!,
\end{eqnarray}
in which all the phases $\phi$ cancel exactly for diagonal entries ($l=l'$). As a result, for a diagonal operation we obtain
\begin{eqnarray}\label{eq:app:pulses:U-final}
U(t_f,t_i) = \sum_{l}B_{ll}\v{a^n_l}\iv{a^n_{l}}.
\end{eqnarray}
Note that Eq.~(\ref{eq:app:pulses:U-final}) can be applied iteratively and, hence, generalized to larger pulse sequences provided all the conditions stated above are met. 

However, if $U_0$ is designed to performs a {\it non-diagonal} quantum operation, e.g., a {\it single}- or {\it two-qubit} swap, the phases of the pulse(s) performing $U_0$ will not cancel out completely.\cite{Yale} The phase difference will appear in the final evolution operator $U(t_f,t_i)$. This can be verified most easily for the example of a {\it single-qubit} swap operation [see Eq.~(\ref{eq:1Q:XY})]. Therefore, pulses performing single-qubit or two-qubit non-diagonal operations have to be phase-locked.

\section{Fidelity of a gate operation}\label{app:fidel}

The fidelity of a gate operating on state $\v{\psi_0}$ is given by
\begin{eqnarray}\label{eq:app:fidel:coherentF}
F(\psi_0,\psi) = |\iv{\psi_0}U_0\v{\psi}|,
\end{eqnarray}
where $U_0$ is the evolution operator corresponding to the
ideal gate, and $\v{\psi} = U\v{\psi_0}$ is the state of the system after the actual gate. The value of $F(\psi_0,\psi)$ depends on the initial
state of the system and therefore can vary depending on the
choice of algorithm and initial data. In order to obtain an estimate suitable for any algorithm, an average fidelity $F$ is evaluated by taking the average over all possible initial
states of the two-qubit system,
\begin{eqnarray}\label{eq:app:fidel:coherentAvF}
F^2&=& \int d\psi_0 F(\psi_0,\psi\{\psi_0\})^2
\\\nonumber
&=&
\sum_{ijnm}
\frac{\delta_{in}\delta_{jm} + \delta_{ij}\delta_{nm}}{20}
\iv{n}U^\dag_0 U\v{i}\iv{j}U^\dag U_0\v{m}.
\end{eqnarray}
The integration $\int d\psi_0$ is performed over all complex
amplitudes that define the initial state.\cite{Pedersena}

In the system open to noise, a separable quantum wave function is no longer accessible, and fidelity has to be defined via the reduced density matrix
\begin{eqnarray}\label{eq:app:fidel:rhoF}
F(\psi_0,\rho\{\psi_0\}) = \sqrt{\iv{\psi_0}U^\dag_0\,\rho\, U_0\v{\psi_0}}.
\end{eqnarray}
In this case the average fidelity is computed as
\begin{eqnarray}\label{eq:app:fidel:rhoF}
F^2\!\! = \!\!\!\!\!\!\!\!\!\!\!\sum_{ijnm=\{1,4\}}
\!\!\!\!\!\!\!\!\!\frac{\delta_{in}\delta_{jm} + \delta_{ij}\delta_{nm}}{20} \iv{n}U^\dag_0 \rho\{\v{i}\iv{j}\}U_0\v{m},
\end{eqnarray}
where $\rho\{\v{i}\iv{j}\}$ is the part of the reduced density matrix obtained as the result of the evolution of $\v{i}\iv{j}$ due to Eq.~(\ref{eq:dec:rho-eq}) or (\ref{eq:dec:rho-E-eq}).
This is the generalization of Eq.~(\ref{eq:app:fidel:coherentAvF}) for the case of non-unitary evolution of a pure initial state. It is possible because the reduced density matrix after the gate operation is a linear function of the initial reduced density matrix.\cite{SolenovFedichkin}

\end{document}